\documentclass[9pt, sigplan, authorversion]{acmart}

\usepackage{amsmath}

\usepackage{booktabs}
\usepackage{outlines}
\usepackage{epsfig}
\usepackage{graphicx}

\PassOptionsToPackage{hyphens}{url}
\usepackage[hyphens]{url}
\usepackage[bottom,para]{footmisc}
\usepackage{multirow}
\usepackage{caption}
\usepackage{xspace} 
\usepackage{lipsum}
\usepackage{subcaption}
\usepackage{amsfonts} 
\DeclareGraphicsExtensions{.png, .pdf, .jpg, .bmp}

\usepackage{filecontents}
\usepackage{tikz}
\usepackage{amsmath}

\usepackage{graphicx}
\usepackage{tabularx}
\usepackage{multirow}
\usepackage{balance}
\usepackage{makecell}
\usepackage{adjustbox}

\usepackage{tcolorbox}

\usepackage{enumitem}
\setlist{topsep=1pt,itemsep=1pt,parsep=1pt,itemindent=0pt,leftmargin=0.13in}
\usepackage{outlines}

\usepackage{soul}
\soulregister\cite7
\soulregister\ref7

\usepackage{changes}

\newcommand{\ie}{{\em i.e., \/}}
\newcommand{\eg}{{\em e.g., \/}}
\newcommand{\etc}{{\em etc. \/}}

\newcommand{\hide}[1] {}

\newcommand{\knote}[1]{{\color{red}[KK: #1]}}

\newcommand{\sknote}[1]{{\color{magenta}[SK: #1]}}

\renewcommand{\knote}[1]{\hide{{\color{red}[KK: #1]}}}
\renewcommand{\sknote}[1]{\hide{{\color{magenta}[SK: #1]}}}

\usepackage{booktabs} 
\usepackage{listings}
\usepackage{color}

\usepackage{url}

\definecolor{dkgreen}{rgb}{0,0.6,0}
\definecolor{gray}{rgb}{0.5,0.5,0.5}
\definecolor{mauve}{rgb}{0.58,0,0.82}

\newcommand{\Scut}[1] {}

\AtBeginDocument{%
  \providecommand\BibTeX{{%
    \normalfont B\kern-0.5em{\scshape i\kern-0.25em b}\kern-0.8em\TeX}}}

\acmYear{2019}\copyrightyear{2019}
\setcopyright{acmcopyright}
\acmConference[WOSC '19]{Fifth International Workshop on Serverless Computing}{December 9--13, 2019}{Davis, CA, USA}
\acmBooktitle{Fifth International Workshop on Serverless Computing (WOSC '19), December 9--13, 2019, Davis, CA, USA}
\acmPrice{15.00}
\acmDOI{10.1145/3366623.3368139}
\acmISBN{978-1-4503-7038-7/19/12}

\begin{document}

\setlength{\pdfpageheight}{\paperheight}
\setlength{\pdfpagewidth}{\paperwidth}

\title{Understanding Open Source Serverless Platforms:\\ Design Considerations and Performance}

\author{Junfeng Li$^{1,2}$, Sameer G. Kulkarni$^2$, K. K. Ramakrishnan$^2$, and Dan Li$^1$}
\affiliation{%
    \institution{$^{1}$Tsinghua University  $^{2}$University of California, Riverside}
}








\renewcommand{\shortauthors}{J. Li, et al.}
\renewcommand{\shorttitle}{Understanding Open Source Serverless Platforms: Design Considerations and Performance}

\vspace{-2mm}
\begin{abstract}
\vspace{-2mm}
Serverless computing is increasingly popular because of the promise of lower cost and the convenience it provides to users who do not need to focus on server management. 
This has resulted in the availability of a number of proprietary and open-source serverless solutions.
We seek to understand how the performance of serverless computing depends on a number of design issues using several popular open-source serverless platforms.
We identify the idiosyncrasies affecting performance (throughput and latency) for different open-source serverless platforms. 
Further, we observe that just having either resource-based (CPU and memory) or workload-based (request per second (RPS) or concurrent requests) auto-scaling is inadequate to address the needs of the serverless platforms.
\end{abstract}\vspace{-2mm}




\maketitle

\vspace{-2mm}
\section{Introduction}\vspace{-2mm}
Serverless computing has ushered in a new era in cloud computing. 
Cloud computing 
seeks to provide compute and storage services at large scale and low cost to end-users through economies of scale and effective multiplexing.    
Serverless computing takes this multiplexing and scalability to the next level by allowing providers to commit just the required amount resources to a particular application 
(as many instances as necessary, but only when needed) and utilize the resources for just the time needed to execute an invoked function. Resources are scaled dynamically to meet the demand from user requests.
Unlike the `traditional' cloud deployment model
, where the number of necessary compute instances are deployed well in advance,
serverless computing allows the cost to be near zero when there is no demand, and scales to as many instances as needed to meet the traffic demand. Thus, serverless  is meant to be both scalable and more cost effective. 

In addition to scaling and multiplexing, serverless computing allows developers to build, deploy and run the application on demand without focusing on server management, according to the Cloud Native Computing Foundation (CNCF)~\cite{yaroncncf}.  
When an event is triggered, a piece of infrastructure is allocated dynamically to execute the code. 
The underlying details of resource management: resource allocation, communication of user data, and the execution of functions is abstracted from the user.
Serverless computing 
manages cloud resources typically by deploying applications in dynamically instantiated containers. For instance, Amazon provides AWS-Lambda~\cite{awslambda}, an event-driven, serverless computing platform that enables to implement and deploy application code in any of the supported languages 
and execute on-demand as docker-containers. The serverless infrastructure manages the queuing of requests and can automatically scale 
containers to meet fluctuating demands.

Our focus is
not only on the evaluation and comparison of performance, but seek to identify the key differences in the workings of different Kubernetes-based open-source serverless platforms.
We systematically identify the strengths and deficiencies of 
Knative\footnote{\url{https://github.com/knative}}, Kubeless\footnote{\url{https://kubeless.io}}, Nuclio\footnote{\url{https://nuclio.io}} and OpenFaaS\footnote{\url{https://www.openfaas.com}}. Our key contributions include:
\begin{itemize}
    \item We provide an understanding of the role and interaction of the different components of each of these platforms. 
    \item We describe the impact of key configuration parameters of different components (platform, gateway, controller and function).
    \item We evaluate the mode and operation of auto-scaling supported by these different platforms for different kinds of workloads.
\end{itemize}
 


\vspace{-2mm}
\Scut{
\begin{figure*}[]\vspace{-3mm}
	\centering
		\centering
	\begin{subfigure}{.475\textwidth}
		\includegraphics[width=\linewidth]{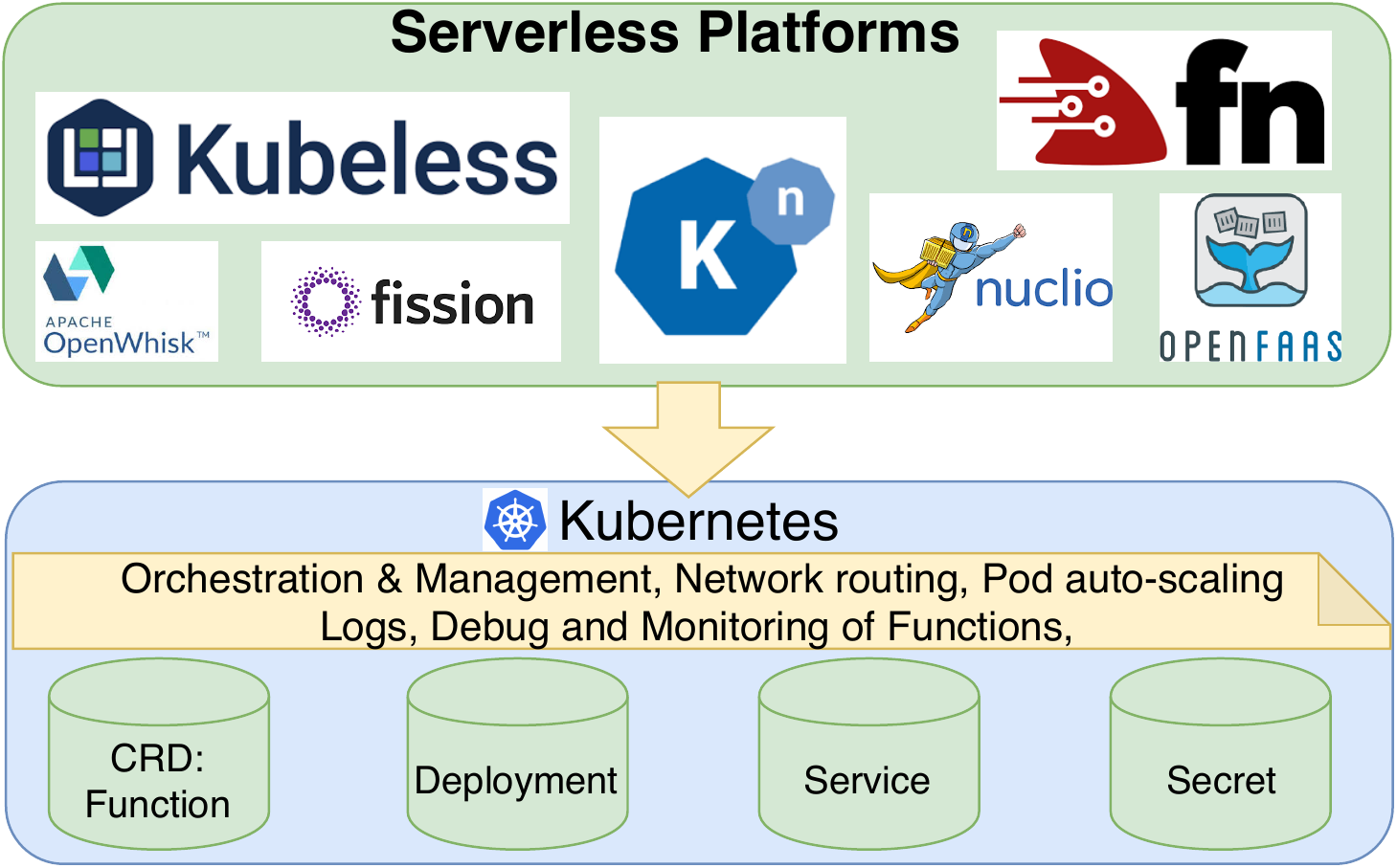}
		\vspace{-5mm}
		\caption{Kubernetes Services used by Serverless Frameworks}
                \label{fig:kubernets_svcs}
	\end{subfigure}%
	\hspace{2mm}
	\begin{subfigure}{.475\textwidth}
	\includegraphics[width=\textwidth]{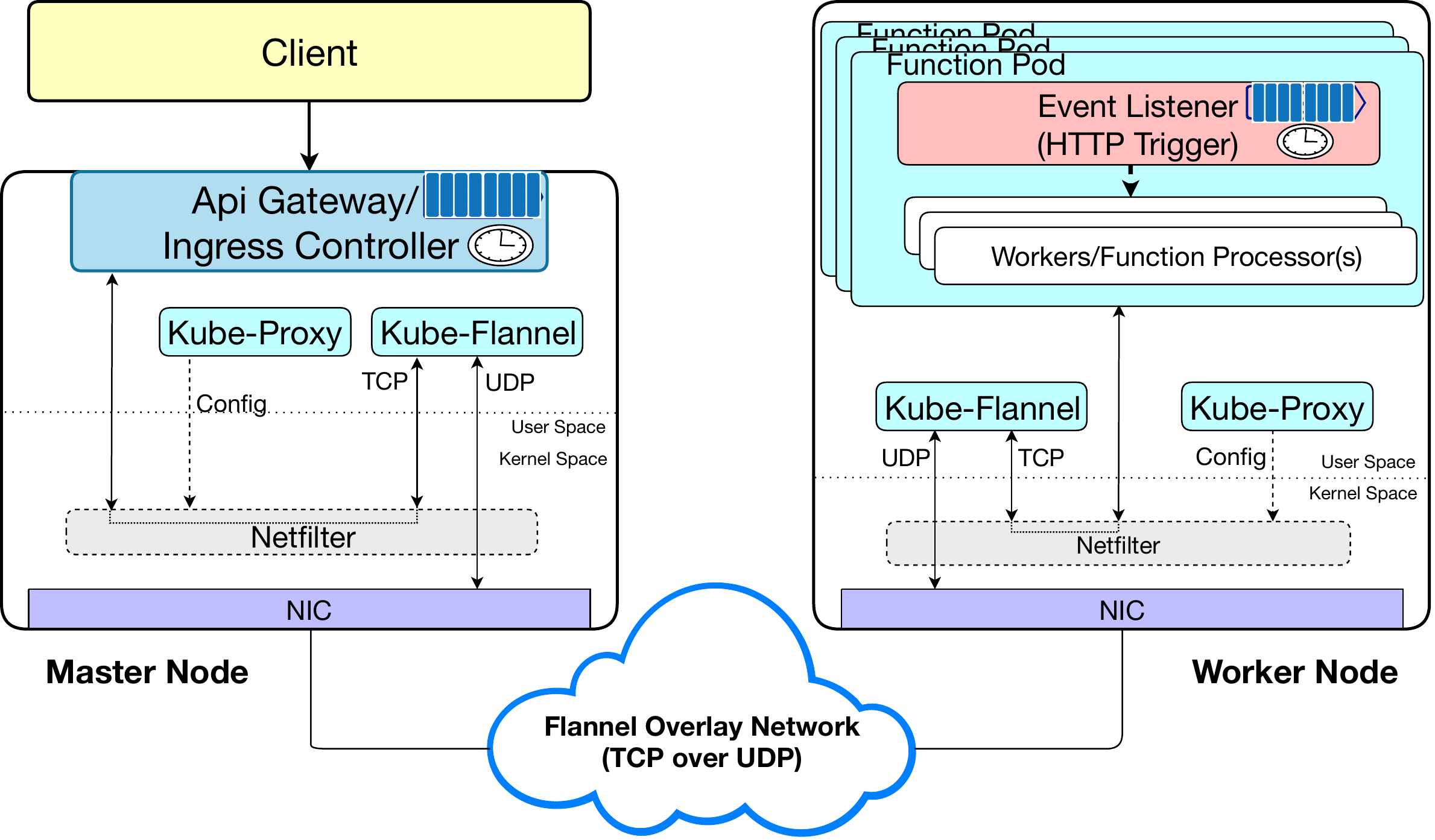}
	\vspace{-5mm}
	\caption{Kubernetes: Network routing (Flannel mode) to export the services.}
    \label{fig:svc_export_flannel_arch}
    \end{subfigure}
\vspace{-3mm}
\caption{Kubernetes architecture.}
\vspace{-4mm}
\end{figure*}
}

\begin{figure}[]
	\centering
	\includegraphics[width=\columnwidth]{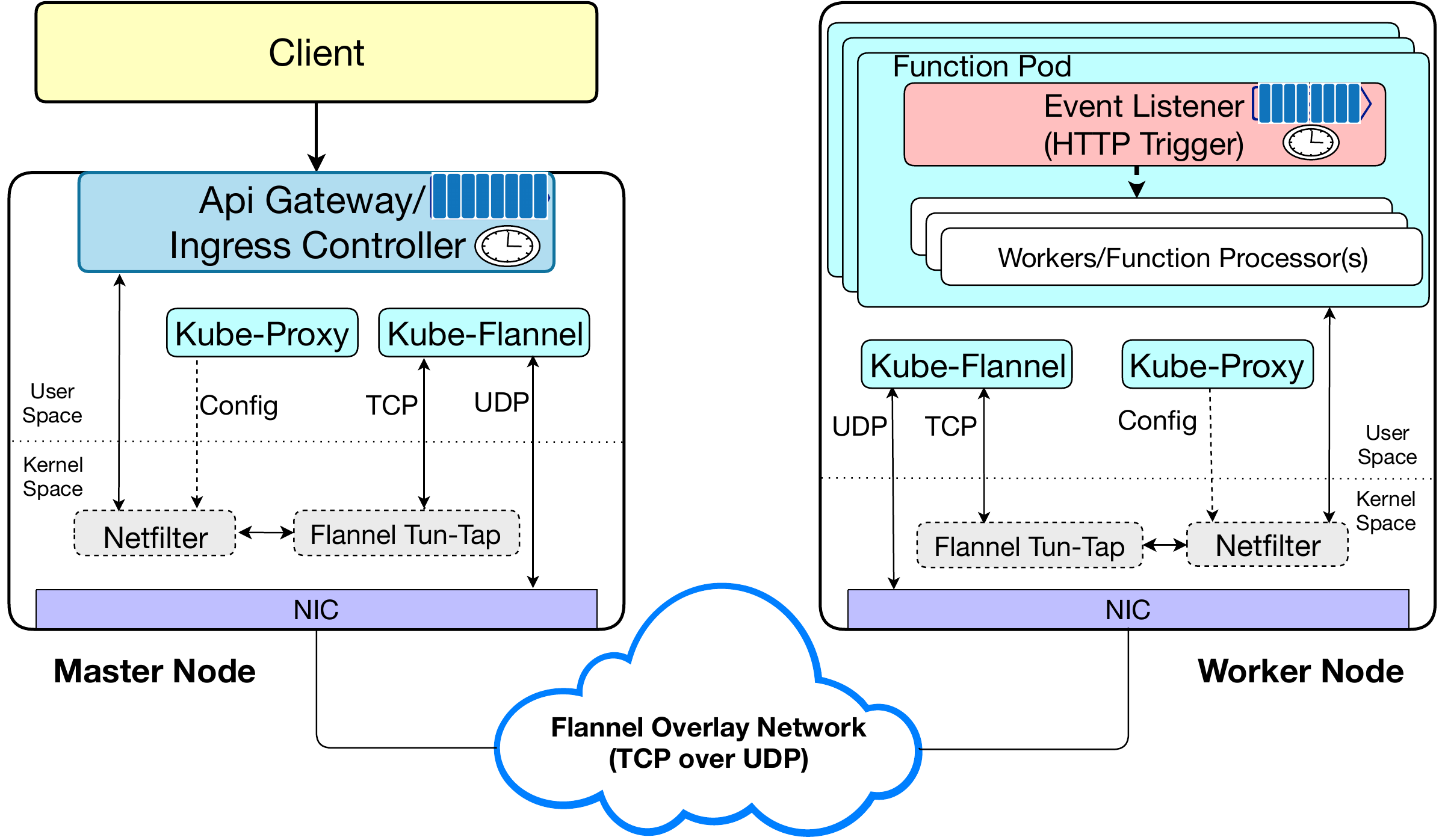}
	\vspace{-8mm}
	\caption{Kubernetes: Network routing to export the services.}
    \label{fig:svc_export_flannel_arch}
\vspace{-4mm}
\end{figure}
\section{Background \& Comparison}\vspace{-2mm}
\begin{figure*}
	\centering
	\begin{subfigure}{.475\textwidth}
		\includegraphics[width=\linewidth]{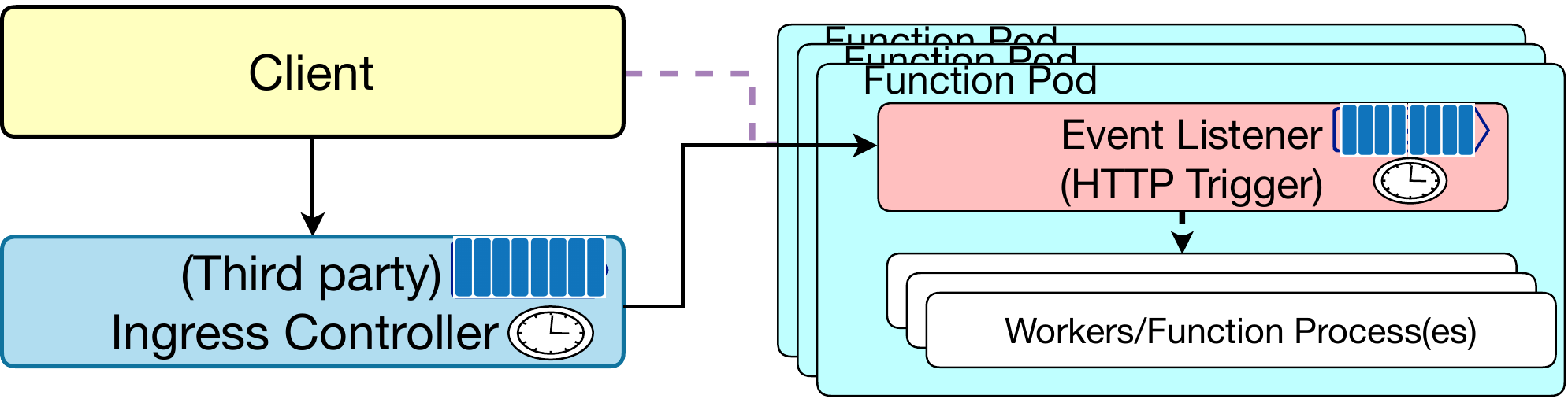}
		\vspace{-5mm}
		\caption{Nuclio Serverless Platform}
                \label{fig:nuclio_arch}
	\end{subfigure}%
	\hspace{2mm}
	\begin{subfigure}{.475\textwidth}
		\includegraphics[width=\linewidth]{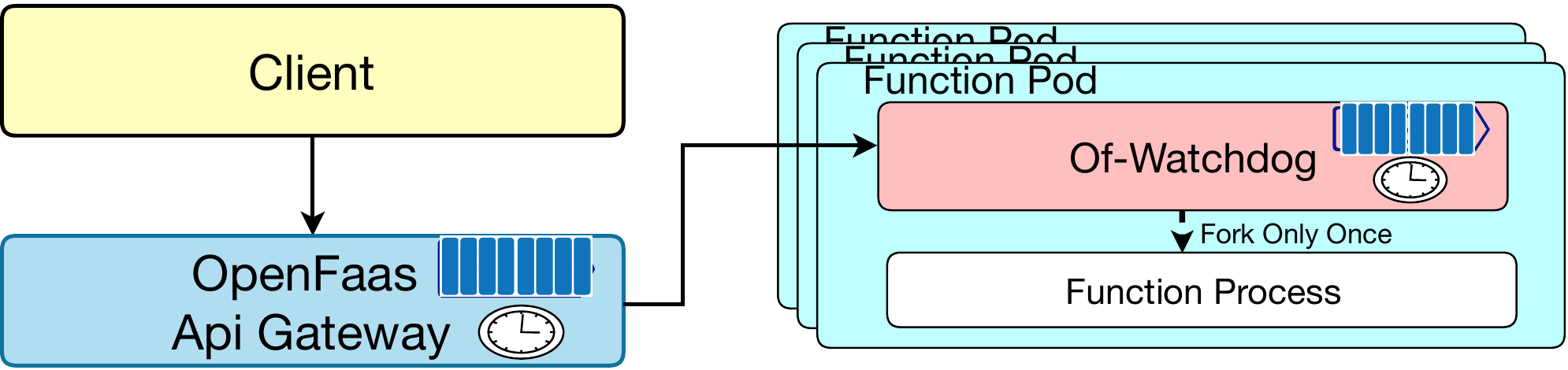}
		\vspace{-5mm}
		\caption{OpenFaas Serverless Platform}
                \label{fig:openfaas_arch}
	\end{subfigure}

	\centering
	\begin{subfigure}{.475\textwidth}
		\includegraphics[width=\linewidth]{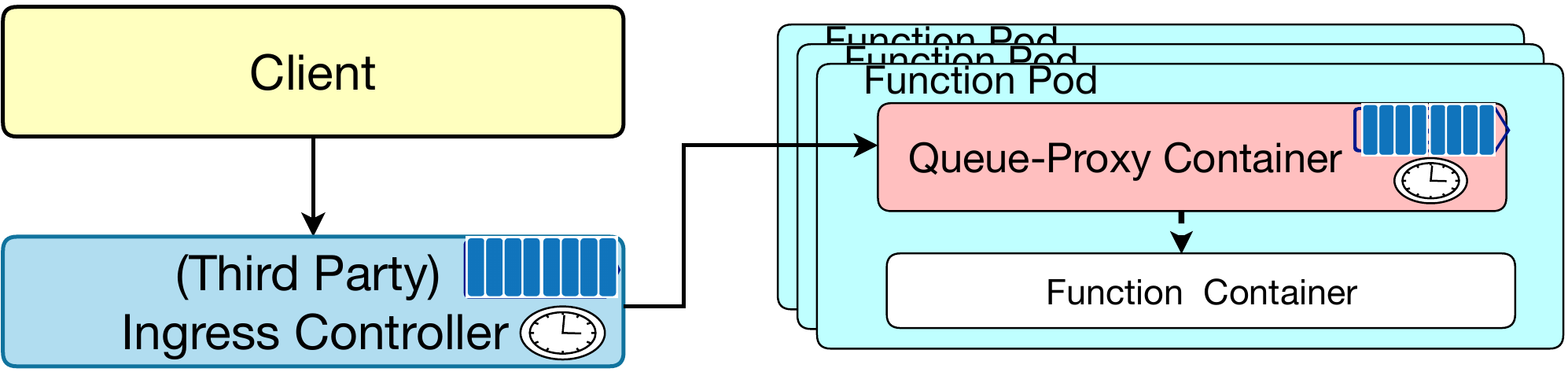}
		\vspace{-5mm}
		\caption{Knative Serverless Platform}
                \label{fig:knative_arch}
	\end{subfigure}%
	\hspace{2mm}
	\begin{subfigure}{.475\textwidth}
		\includegraphics[width=\linewidth]{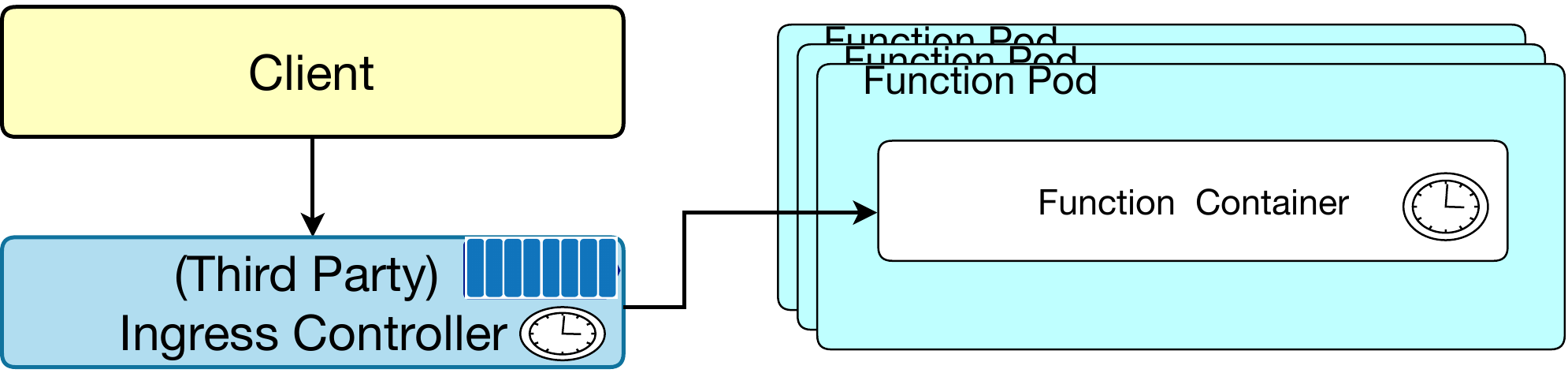}
		\vspace{-5mm}
		\caption{Kubeless Serverless Platform}
                \label{fig:kubeless_arch}
	\end{subfigure}
	\vspace{-4mm}
	\caption{Working model for different kubernetes-based serverless platforms (Nuclio, OpenFaas, Knative and Kubeless).}\label{fig:serverless_platforms_arch}
	\vspace{-4mm}
\end{figure*}



Several cloud service provides (CSPs) offer serverless computing platforms on their public clouds \eg AWS Lambda functions, Google Cloud Platform, Microsoft Azure, and IBM Bluemix \etc These cloud platforms also offer other supporting services such as an event notification service, storage service, database services \etc that are necessary for operating an overall serverless ecosystem. 
These CSPs govern many of the function-related characteristics such as: how long functions can run, how long can they be kept 
idle, the number of concurrent active instances, load balancing among the active instances, retry in the case of failed requests \etc These are
almost entirely dependent on the cloud providers' terms and conditions. To understand the impact of these choices, it is useful to study the functioning of
open source serverless platforms such as Knative, Kubeless, Nuclio, OpenFaaS, OpenWhisk\footnote{\url{https://openwhisk.apache.org}}, \etc

\vspace{-1mm}\subsection{Open-source Serverless platforms}\vspace{-1mm}
Several open source serverless platforms allow us to freely leverage and mix-and-match different open source services, and to deploy and manage the functions on self-hosted clouds. However, the challenges are the i) readiness (requires learning and setup expertise) of the necessary infrastructure and integration of different services, ii) challenges with management and maintenance of the necessary service infrastructure. iii) lack of technical support. 
Hence, in this work we specifically select four of the Kubernetes~\cite{burns2016borg} based open source serverless frameworks based on the recent popularity,\footnote{Until the release of Knative and Nuclio, the Kubeless and OpenFaaS were shown to be the top two leading serverless platforms in terms of current and planned usage~\cite{newstack_serverless_survey_18}} community support and feature richness of these platforms.
\vspace{-1mm}\subsection{Dependency on Kubernetes}\vspace{-1mm}
Kubernetes is a portable and extensible platform that facilitates both declarative configuration and automation of deployment and management of containerized workloads. The serverless frameworks rely on Kubernetes APIs for orchestration and management of the serverless functions. \Scut{Figure~\ref{fig:kubernets_svcs} shows the key functions and services of Kubernetes that the serverless frameworks depend on. They}Serverless platforms typically extend and provide the Custom Resource Definition (CRD) features necessary to create and deploy the container pods (group of containers). They depend primarily on Kubernetes for i) Configuration management of containers and pods; ii) Pod scheduling and service discovery; iii) Update roll-outs for functions; and iv) Replication management.

\vspace{-1mm}\subsection{Salient Characteristics of Serverless Platforms}\vspace{-1mm}
Fig.~\ref{fig:serverless_platforms_arch} shows the framework and key components of the 4 different serverless platforms considered in this work.
\Scut{We briefly discuss their salient features, their distinct configuration parameters and how their settings influence baseline performance. 
}
\\\noindent{\bf Nuclio}:
\Scut{Nuclio is both an open source and managed serverless platform supported by Iguazio Systems.\footnote{The opensource version does not support the auto-scaling feature.} 
}
Fig.~\ref{fig:nuclio_arch} shows the key components of Nuclio.
The distinct feature of Nuclio is the `Processor' architecture which provides work parallelism through multiple worker processes that can run in each container. 
First, the Nuclio service model supports invocation of the `function' pod directly from an external client, without the need for any ingress controller or API gateway. 
Second, the function pod consists of two kinds of processes namely the i) event-listener and ii) one or more worker (user deployed function) processes.
Note, the event-listener can be configured with a timeout parameter to control how long events can be queued\Scut{ there before expiring, while the requests are waiting for processing. Thus, 
the process can queue multiple requests for processing}. 
Third, the number of worker processes can be setup as a static configuration parameter. This enables the function pod to run a desired amount of function instances as different processes, and allows parallel execution on a multi-core node.  

\begin{figure}[htbp!]\vspace{-2mm}
\begin{subfigure}{.5\columnwidth}
    \centering
    \includegraphics[width=\linewidth]{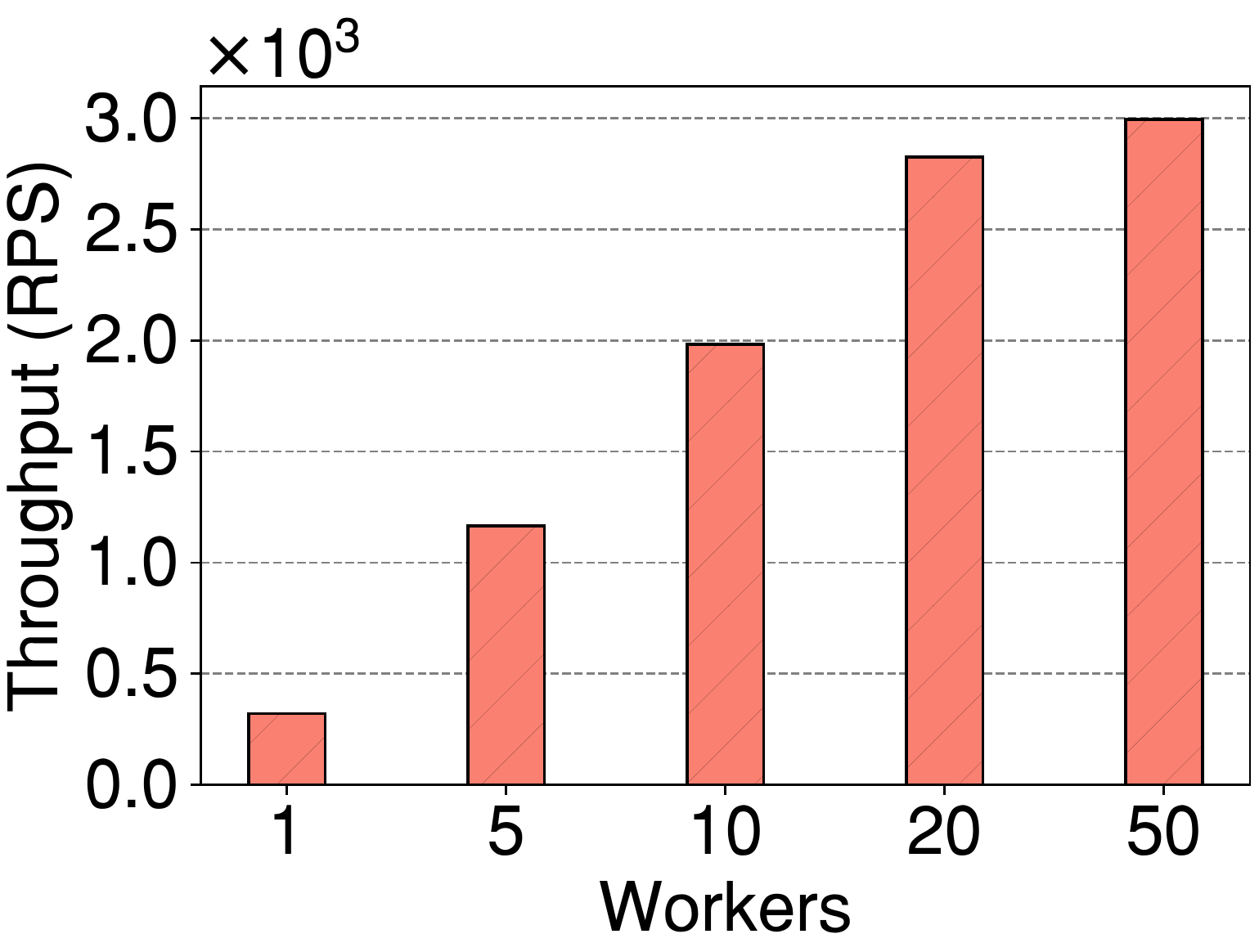}
    \vspace{-6mm}
    \caption{Throughput in requests/second.}
    \label{fig:nuclio_worker_perf}
\end{subfigure}%
\begin{subfigure}{.5\columnwidth}
    \centering
    \vspace{1mm}
    \includegraphics[width=\linewidth]{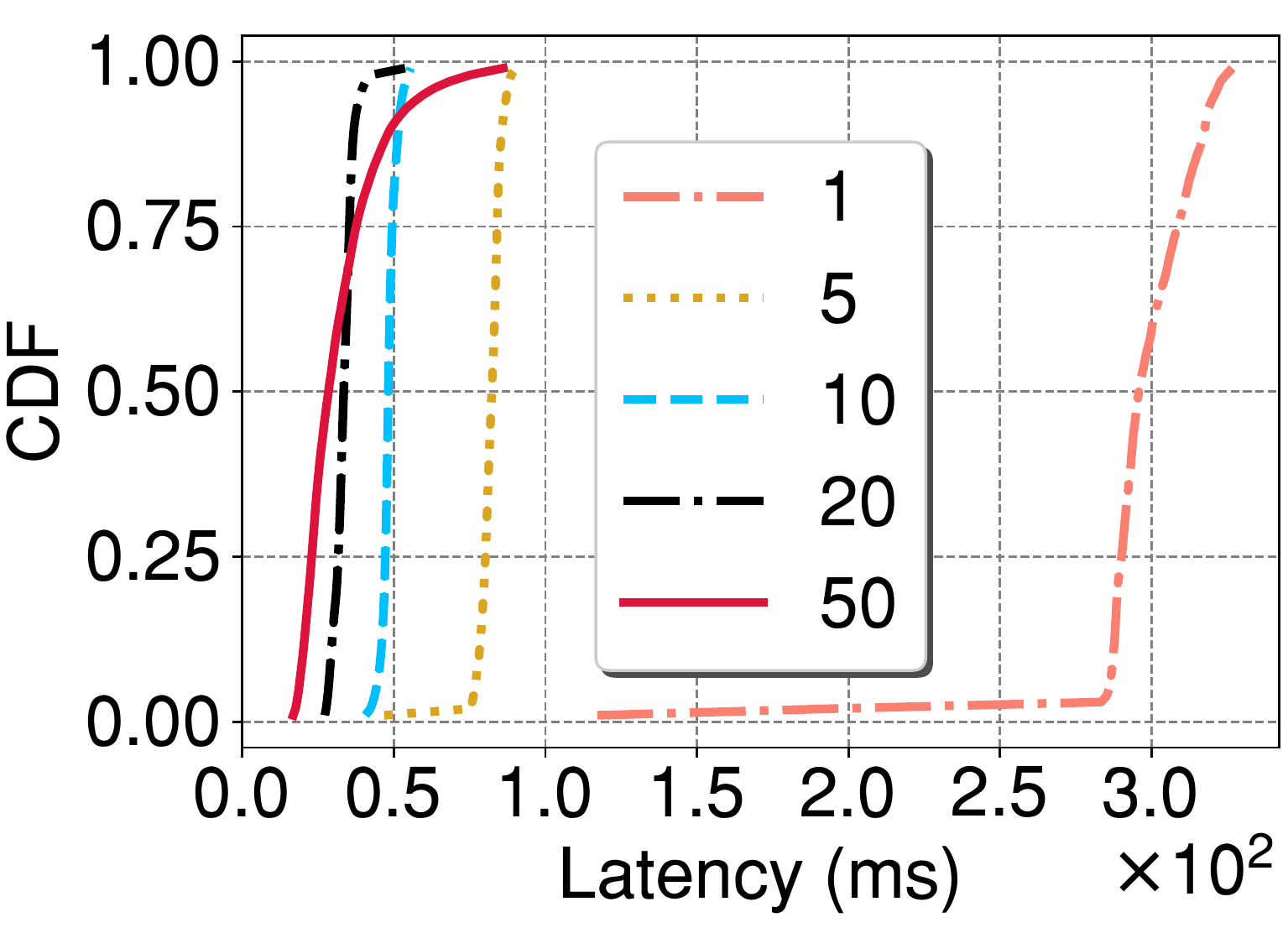}
    \vspace{-6mm}
    \caption{Latency in ms.}
    \label{fig:nuclio_worker_lat}
\end{subfigure}%
\vspace{-4mm}
\caption{Throughput and latency for different number of workers within one Nuclio function pod (100 concurrent requests).}
\label{fig:nuclio_worker}
\vspace{-5mm}
\end{figure}
To quantify the benefit of having multiple workers, we experimented with simple `http-workload' where  we  implement a  simple  python  function  that  communicates  with  a  local  HTTP server (located on Kubernetes master node), to fetch and respond with a 4 byte payload for each of the requests. 
Fig.~\ref{fig:nuclio_worker_perf} shows the impact on throughput and latency for multiple workers. We observe a $4\times$ throughput increase with 4 workers and almost $10\times$ improvement with 50 workers. Note, scaling the number of workers also improves the latency as shown in Fig.~\ref{fig:nuclio_worker_lat}. \Scut{Scaling more workers than the physical cores however only marginally improves the throughput. But more importantly, we find that it adversely impacts the tail latency, because of queuing effects.}
\Scut{However, scaling more workers than physical cores adversely impacts the tail latency due to queuing effect.}

\noindent{\bf OpenFaaS}:
The 
key components of OpenFaaS are shown in Figure~\ref{fig:openfaas_arch}. The API gateway provides access to the functions from outside the Kubernetes cluster (external 
routing), collects metrics and provides scaling by interacting with the Kubernetes orchestration engine.
The API gateway can be scaled to multiple instances.
Also, it can be replaced by a third-party Ingress controller.

\begin{figure}[htb!]\vspace{-3mm}
\begin{subfigure}{.5\columnwidth}
    \centering
    \includegraphics[width=\linewidth]{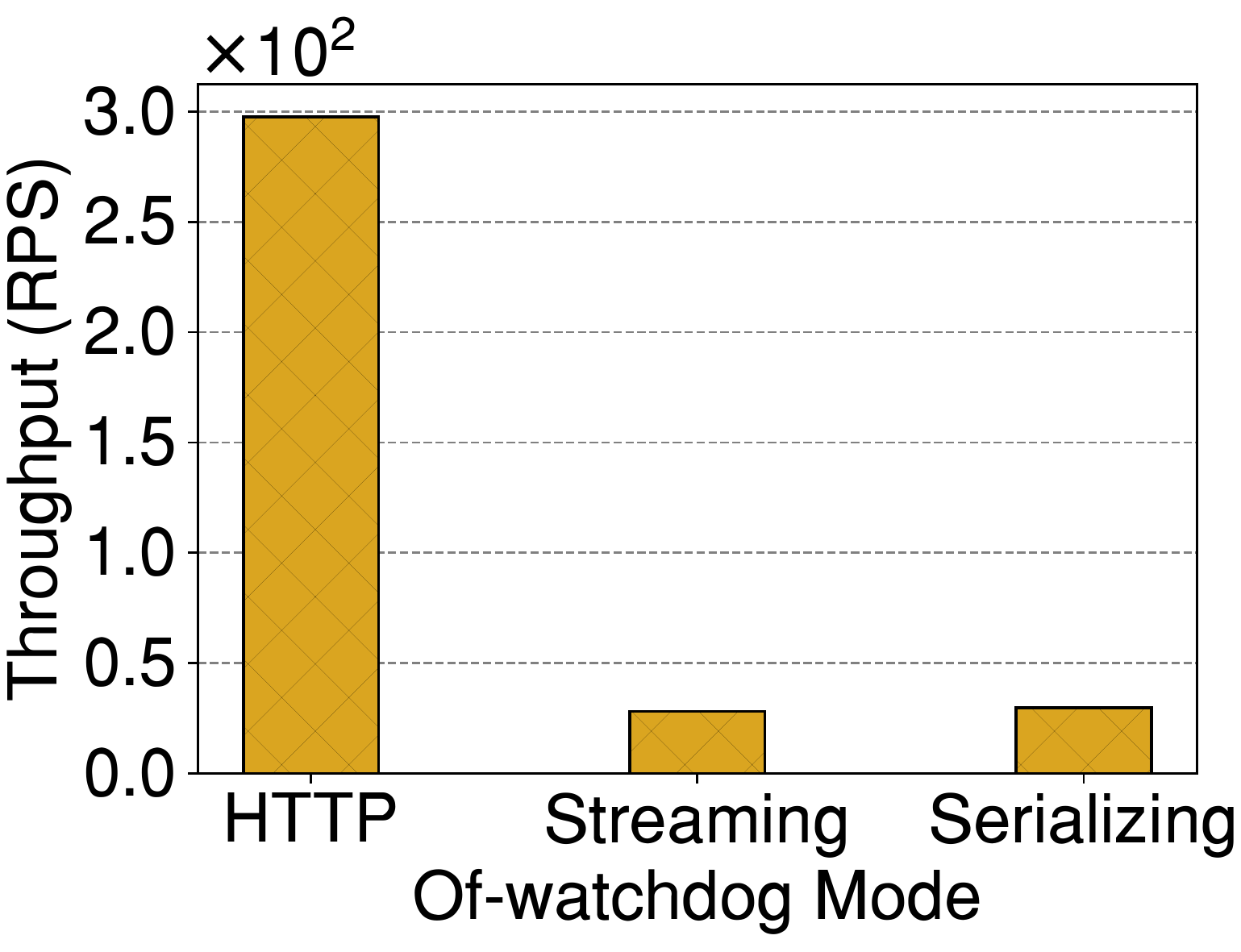}
    \vspace{-5mm}
    \caption{Throughput in requests/second.}
    \label{fig:of_watchdog_perf_rps}
\end{subfigure}%
\begin{subfigure}{.5\columnwidth}
    \centering
    \vspace{1mm}
    \includegraphics[width=\linewidth]{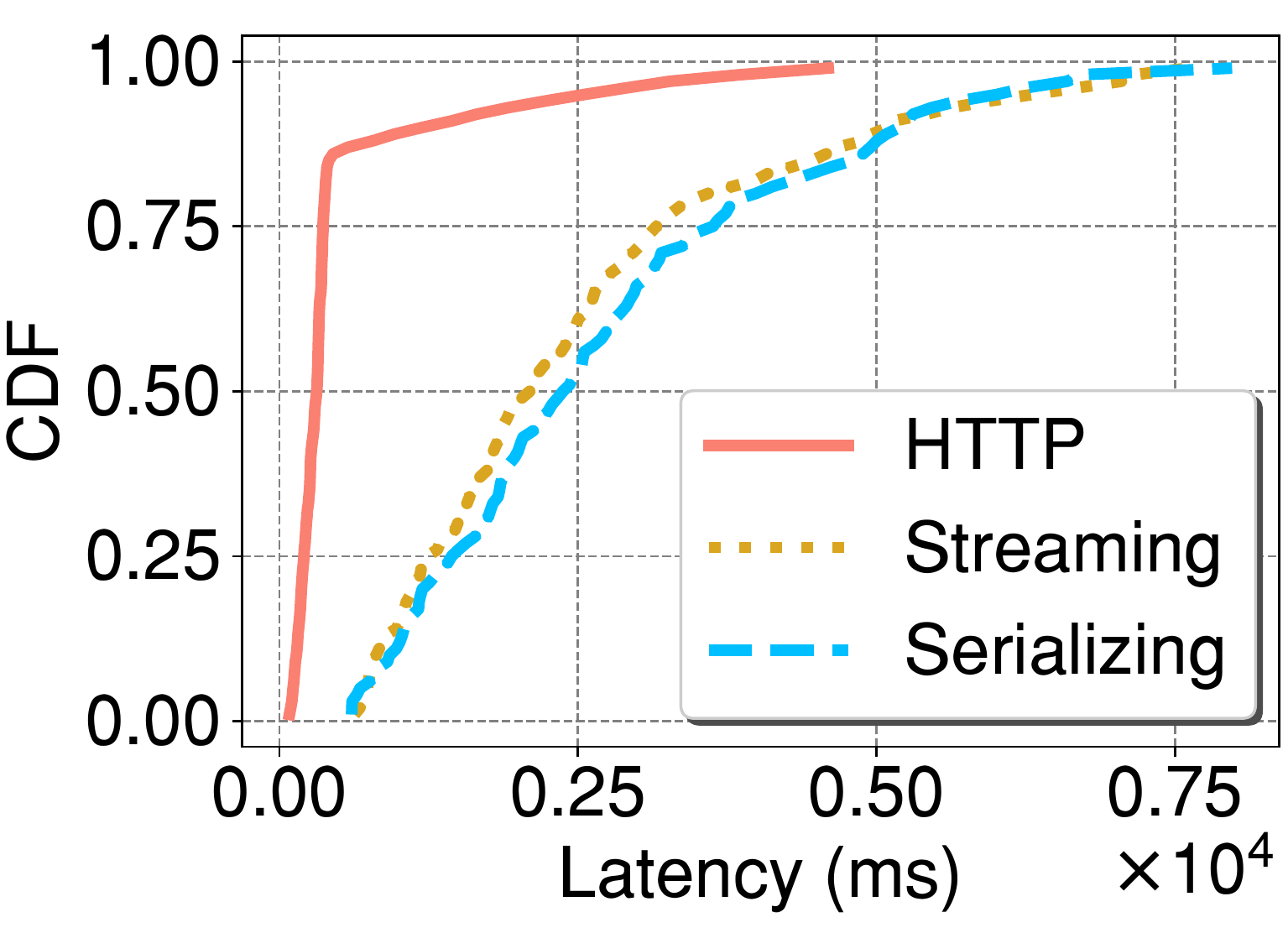}
    \vspace{-5mm}
    \caption{Latency in ms.}
    \label{fig:of_watchdog_perf_latency}
\end{subfigure}%
\vspace{-4mm}
\caption{Throughput and latency for different modes of OpenFaaS of-watchdog (100 concurrent requests).}
\label{fig:of_watchdog_perf}
\vspace{-4mm}
\end{figure}

Each function pod consists of a single container running two kinds of processes namely the i) `of-watchdog' and ii) user deployed function process. The `of-watchdog' is a tiny Golang HTTP server that serves as the entry-point for HTTP requests to be forwarded to the function process. 
Based on use case requirements, the `of-watchdog' can be operated in 3 modes, i.e., `HTTP', `streaming' and `serializing'. In `HTTP' mode, the function is forked only once to one instance (worker) at the beginning and kept warm for the entire life-cycle of the function pod. In both the `streaming' and `serializing' mode, a new function instance (worker) is forked for every request, resulting in significant cold-start latency and impact on the throughput.
Fig.~\ref{fig:of_watchdog_perf} shows the throughput and latency when running the watchdog in different supported modes. The `streaming' mode results in very low performance and is typically only desirable for memory-heavy workloads, while the `serializing' mode is equally poor due to fork per request. For our subsequent evaluation, we choose the `HTTP' mode.

\noindent{\bf Knative}:
Fig.~\ref{fig:knative_arch}, shows the key components of Knative.
We see that each function pod consists of two containers namely the `queue-proxy' and the `function'. The `queue-proxy' is responsible for queuing incoming requests and forwarding them to the `function' container for execution. It also handles the timeout of queued requests. This queue enables the worker to quickly fetch requests from the ingress controller and process them, thus achieving better throughput, although incurring queuing latency.
Interestingly, we can observe that since Knative implements the `queue-proxy' and `function' as two different containers in a pod, the communication overhead is higher than the process model of the Nuclio and OpenFaaS, resulting in relatively lower performance. 
\begin{figure}[htb!]\vspace{-4mm}
\begin{subfigure}{.5\columnwidth}
    \centering
    \includegraphics[width=\linewidth]{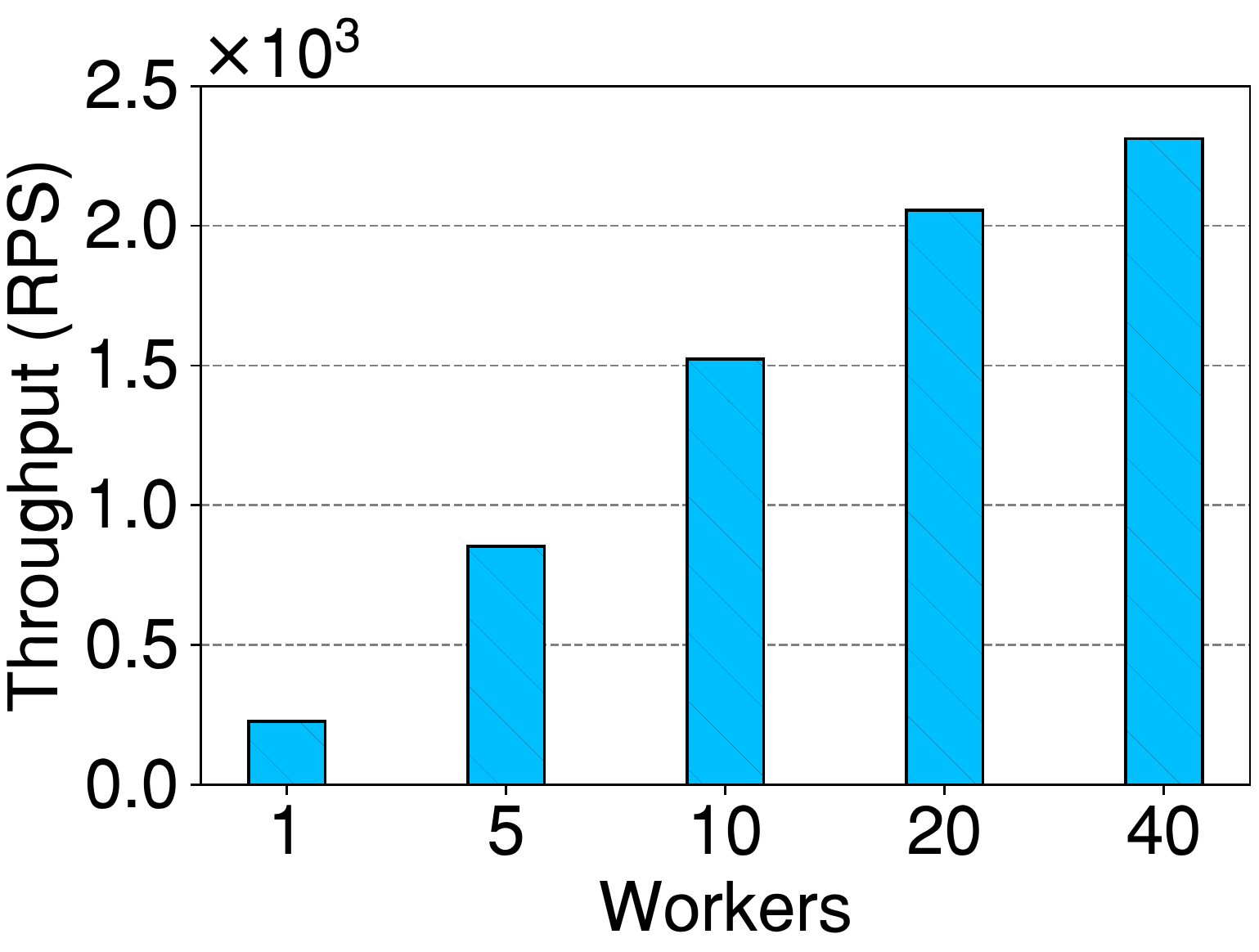}
    \vspace{-6mm}
    \caption{Throughput in requests/second.}
    \label{fig:knative_worker_perf}
\end{subfigure}%
\begin{subfigure}{.5\columnwidth}
    \centering
    \vspace{1mm}
    \includegraphics[width=\linewidth]{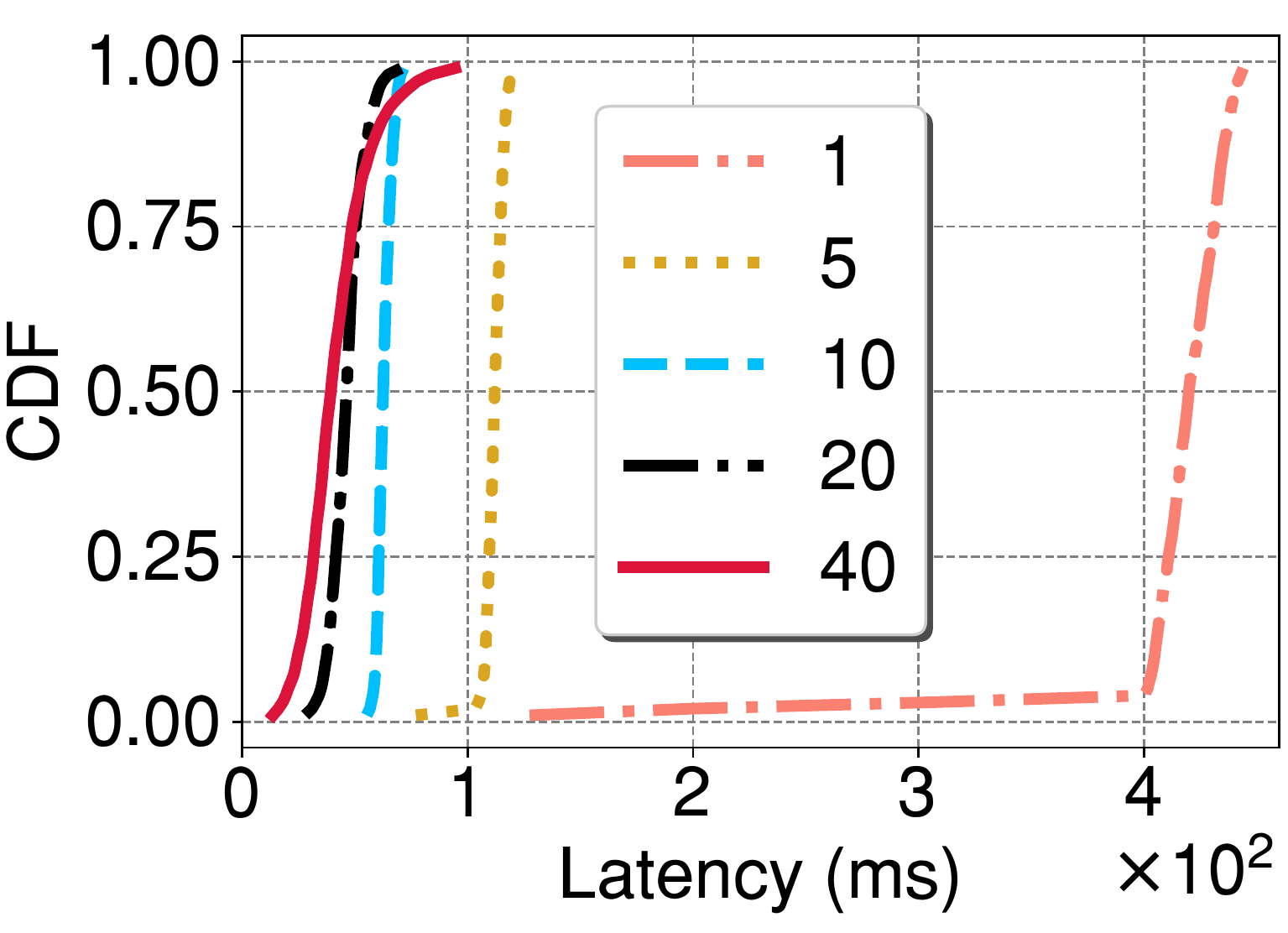}
    \vspace{-6mm}
    \caption{Latency in ms.}
    \label{fig:knative_worker_lat}
\end{subfigure}%
\vspace{-4mm}
\caption{Throughput and latency for different number of workers within one Knative function pod (100 concurrent requests).}
\vspace{-4mm}
\label{fig:knative_worker}
\end{figure}

\vspace{1mm}
For the python runtime, we observed that Knative levarages gunicorn\footnote{\url{http://gunicorn.org}} - a Python web server gateway interface (WSGI) server, 
which supports the pre-fork worker model to create multiple worker processes in a function pod. However, unlike Nuclio, the number of concurrent workers is not exported as a configuration parameter for deployment. 
Figure~\ref{fig:knative_worker} shows the impact on throughput and latency for multiple workers. The characteristics are similar to the Nuclio workers, discussed earlier.

Another distinct feature of Knative is the `panic mode' scaling mechanism of the autoscaler component. 
Panic mode enables the autoscaler to be more responsive to sudden traffic spikes (two times the desired average traffic or a configured threshold value) by quickly scaling the functions instances (up to $10\times$ the current pod count or the maximum configured limit).

\noindent{\bf Kubeless}:
Kubeless is another open source platform built on top of Kubernetes\Scut{, based on the three basic primitives: functions, triggers and runtime. 
A function is a representation of the code to be executed, and trigger is an event source. 
A trigger can be associated with a single function or to a group of functions depending on the event source type.
A runtime represents a language and runtime specific environment in which a function will be executed}.
Figure~\ref{fig:kubeless_arch} describes the key components and the working model of serverless functions in Kubeless. 
We also experimented with NGINX ingress controller,\footnote{\url{https://kubernetes.github.io/ingress-nginx}} and 
opted for Traefik due to better performance.


\vspace{-1mm}\subsubsection{Exporting Services and Network routing}\vspace{-1mm}
Serverless frameworks leverage Kubernetes network model to export services (cluster of function pods) and to route requests to specific functions. An API Gateway and/or the Ingress Controller components of the serverless platform can be either exported with a public IP address or can also use the Kubernetes networking model to export the services. 
Fig.~\ref{fig:svc_export_flannel_arch} describes `Flannel' - a simple Kubernetes overlay networking framework to export serverless functions and route the traffic to function pods. 
The Kube-Proxy component of Kubernetes is responsible for setting up the routing and load-balancing rules (\eg setup the netfilter rules to intercept network packets and change their destination/routing) of the traffic intended for Kubernetes pods, while the Kube-Flannel pod is responsible for intercepting the packets destined for Kubernetes pods (listen to traffic for the virtual Kubernetes pod IP range) and performing UDP encapsulation/decapsulation for the traffic exiting/entering the physical network interface.  
In Fig.~\ref{fig:svc_export_flannel_arch}, once the API Gateway/Ingress Controller receives client packets, and determines the service (function) to be executed, it leverages the Kube-Proxy and Kube-Flannel to load-balance and route the traffic to a specific function pod of a worker node. With Kube-Flannel, the traffic leaving the physical network interface is encapsulated and carried over an unreliable UDP transport.

\noindent \textbf{Impact of Ingress Controller and API Gateway components:}
Typically, the API Gateway components enable the URL based routing to different services in a Kubernetes cluster. The function pods are dynamic entities that can be created and destroyed any time because of zero-scaling, auto-scaling, failures \etc Hence, Kubernetes provides service (a virtual cluster with fixed IP, \textsl{a.k.a.} `Cluster IP') as an abstraction to access the pods of a similar kind. 
The API Gateway/Ingress controllers can route the incoming requests in two possible ways: i) route the incoming traffic to the service and let Kubernetes control load-balancing of the traffic across active pods (\eg with the OpenFaaS API Gateway); ii) load-balance and route the traffic directly to any of the active pod instances (\eg with the Knative-Istio ingress controller).


In the former case (API Gateway), we observed that, in order to avoid the overhead of connection setup time, the API Gateway (OpenFaaS API Gateway) sets up multiple connections with the service `Cluster IP'\footnote{Service being a logical entity, the actual TCP connections are setup with different active pods based on the 
Kubernetes routing/load-balancing rules (\eg netfilter rules).} at the beginning (the first access to the function) and it uses these connections to forward 
subsequent requests. No new connections are setup afterwards, unless the existing connections get terminated.
Note that if the connections are not setup after auto-scaling, the traffic cannot get distributed to the newly created pods, thus significantly impacting the performance with auto-scaling (refer \S\ref{eval:autoscaling}). 
However, in the second case (case ii), the ingress controller needs to keep track of the health and status of all the active pods and setup the connections explicitly with each of the active pods to load-balance the traffic.


\vspace{-2mm}
\section{Evaluation}\vspace{-1mm}
The main focus of our evaluation is to distinguish and illustrate the impact of the serverless platform specific design choices and their dependency on the Kubernetes orchestration and management services. A second important focus is to understand the auto-scaling capabilities, and the need to go beyond the resource utilization based scaling services provided by Kubernetes. 

\vspace{-1mm}\subsection{Experimental setup and Workload description}\vspace{-1mm}
We evaluate the serverless platforms on the Cloudlab testbed~\cite{cloudlab} consisting of one master and two worker nodes, each of them equipped with Intel CPU E5-2640v4@2.4GHz (10 physical cores), running Ubuntu 16.04.1 LTS (kernel 4.4.0-154-generic).
We built all four serverless platforms on Kubernetes (v1.15.3), using the latest version available at the time of writing.\footnote{Nuclio (v1.1.16). OpenFaaS consists of: Gateway (v0.17.0), Faas-netes (v0.8.6), Prometheus (v2.11.0), Alert manager (v0.18.0), Queue worker (v0.8.0) and Faas-cli (v0.9.2), and the HTTP mode of-watchdog. 
We use Knative (v0.8) with Istio (v1.1.7) ingress controller, and Kubeless (v1.0.4) with Traefik ingress controller (v1.7).}
We choose Python 3.6 and implement different serverless functions viz. i) simple Hello-world function as the 
baseline, and ii) HTTP server function that fetches and serves pages of different sizes from the local HTTP server.
We use 
\texttt{wrk}~\cite{wrk} 
to generate the HTTP workloads and invoke the serverless functions. \Scut{We detail the platform and workload specific configurations in each of the experiments.} 

\vspace{-1mm}\subsection{Performance - Throughput and Latency}
\vspace{-1mm}\subsubsection{Baseline Performance}~\label{sec:basline_eval}
To evaluate the baseline performance \ie throughput (average requests processed per seconds) and response latency of different serverless platforms, we use a simple `Hello-world' - a no operation function, that  returns 4 bytes of static text in the response. 
For a fair comparison, we limit to 
a single instance of the function pod, disable auto-scaling and configured the same queue size and timeout parameters (50K requests, and 10s timeout) at the ingress/gateway and function pod components across all the platforms. For Nuclio, we further restricted it to a single worker process.

\begin{figure}[htb!]\vspace{-0mm}
\begin{subfigure}{0.5\columnwidth}\vspace{-3mm}
    \centering
    \includegraphics[width=\linewidth, trim=0.01cm 0.01cm 0.01cm 0.01cm, clip=true]{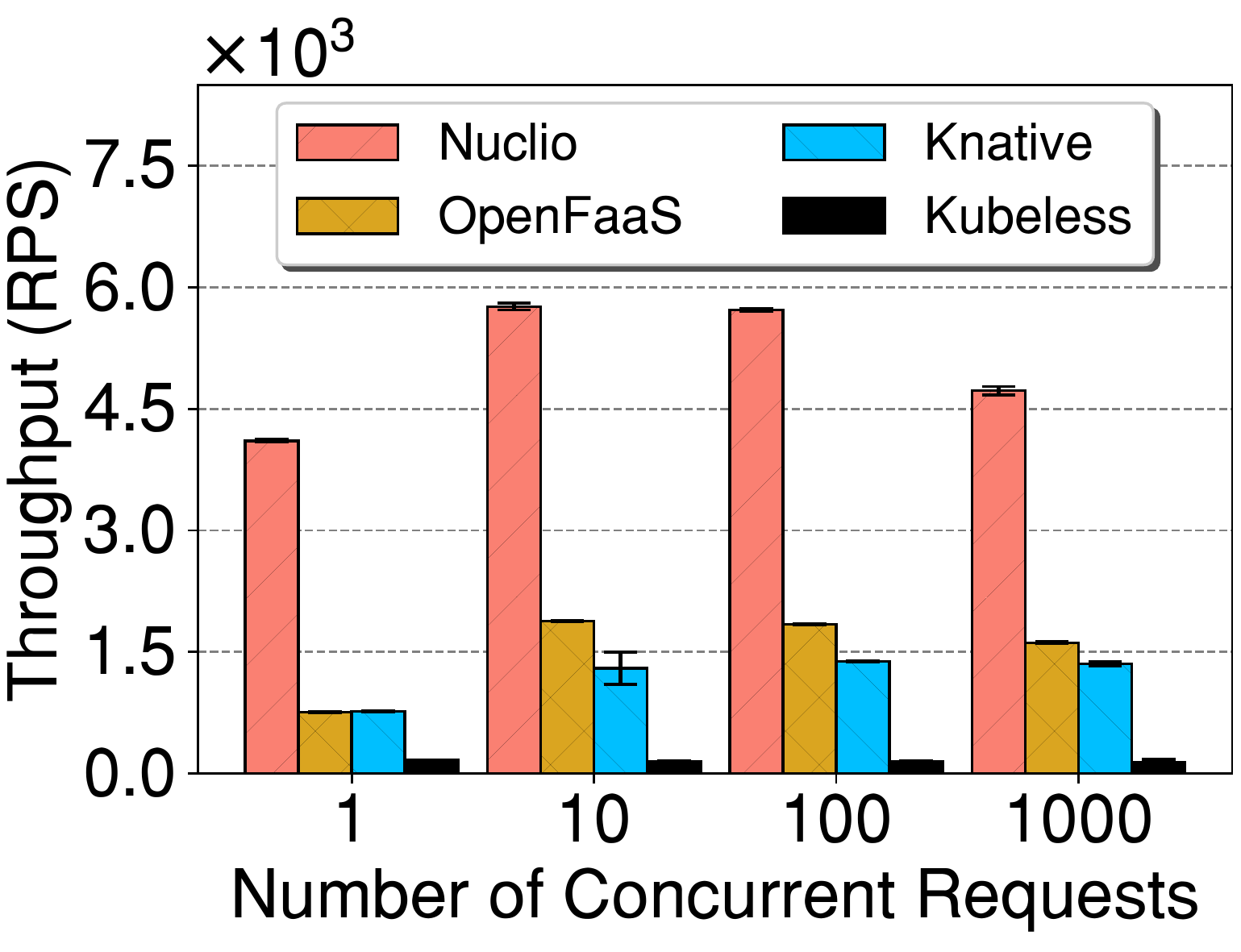}
    \vspace{-5mm}
    \caption{Avg. throughput.}\Scut{ at different concurrency levels.}
    \label{fig:nullthrput}
\end{subfigure}%
\begin{subfigure}{0.5\columnwidth}\vspace{-3mm}
	\centering
	\vspace{1mm}
    \includegraphics[width=\linewidth]{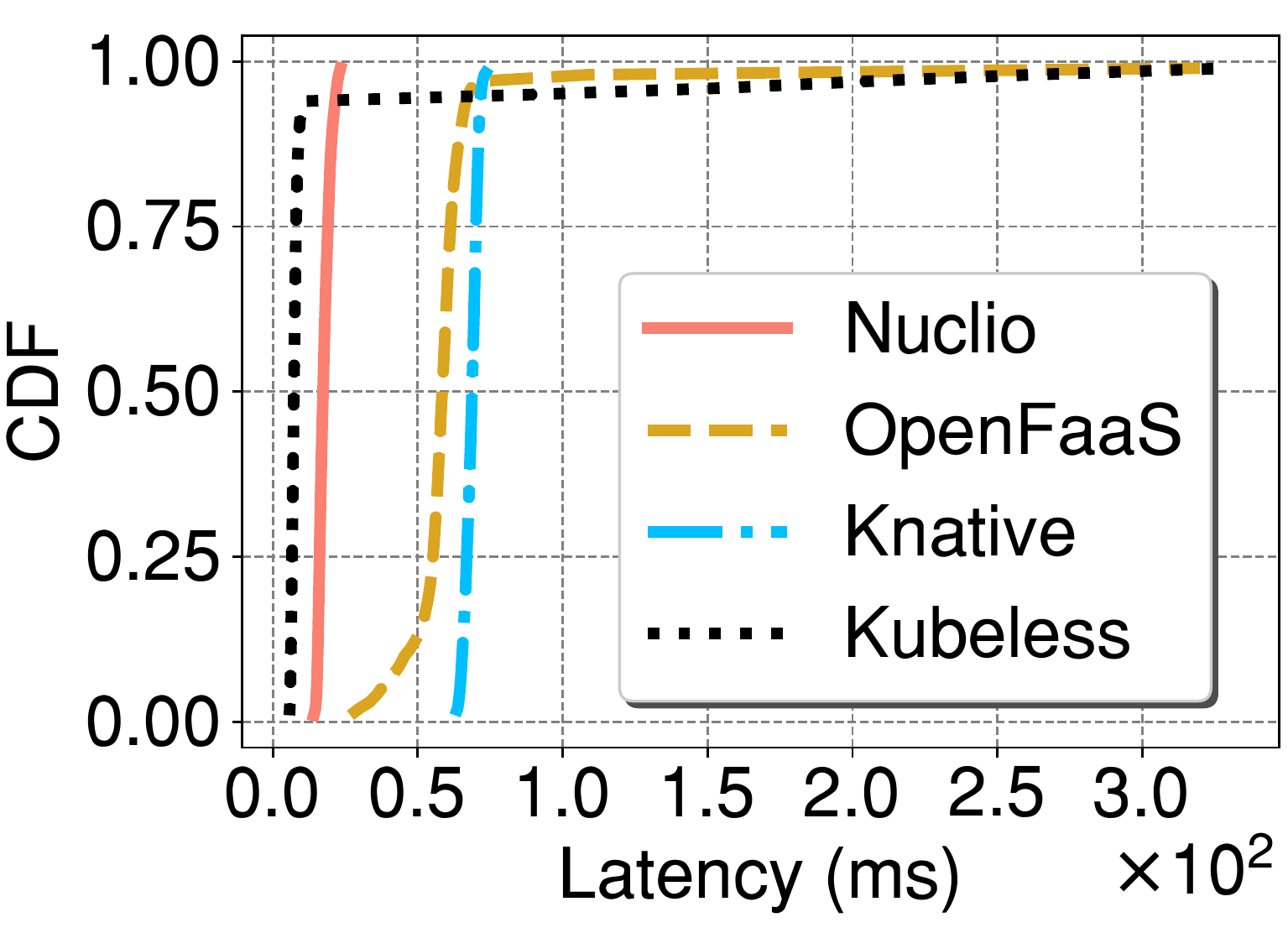}
    \vspace{-5mm}
    \caption{\Scut{CDF of l}Latency for concurrency of 100.}
    \label{fig:nulllatency}
\end{subfigure}%
\vspace{-4mm}
\caption{Throughput and latency of `hello-world' function. 
}\label{fig:NULLThroughput}
\vspace{-4mm}
\end{figure}

Fig. \ref{fig:nullthrput} shows the baseline throughput achieved by different platforms for different concurrent executions of requests. Nuclio outperforms the other platforms due to the low overhead of a direct function call. Routing through the API gateway/Ingress controller components incurs not just the overhead for HTTP connection termination, but also for the context-switch/transfer of the packets across the kernel and user-space of the worker node twice to get it routed to the function pod as shown in the Fig.~\ref{fig:svc_export_flannel_arch}. 
To quantify the overhead, we also experimented with Nuclio using an ingress controller mode and observed the overhead. It resulted in almost half the throughput ($\sim$1700 RPS as opposed to $\sim$3000 RPS for direct call) and nearly $2\times$ latency overhead (increases from 356$\mu$s to 611$\mu$s). 
At the other extreme, Kubeless forks the function for every request, resulting in severely degraded throughput and latency.

From Fig.~\ref{fig:nulllatency}, we can observe that median latency is lowest for Kubeless, and is marginally higher (20$\sim$50 ms) for the queue based frameworks. However, tail latency (above 95\%ile) degrades severely for Kubeless 
and OpenFaaS, while Nuclio and Knative do not see this increased heavy tail-latency. \knote{this sentence is not understandable. Also, call out the two latency graphs as (b) and (c), explain why you have them and then discuss the latency. What you are implying is that the tail latency for Knative is high, right? What do you mean by queue-based? Do you mean concurrency level is 1000?}\sknote{ we wanted to show latency with concurrency=1 (no queuing) and concurrency=1000 (queuing effect) differences, but due to space limits, and small plots, we settled to show only for concurrency=100.} 
The results indicate that having process based communication within a container (\eg Nuclio) along with a local worker queue 
achieves better throughput by having lower overhead for processing requests.


\vspace{-1mm}\subsubsection{HTTP Workload}\label{sec:httpworkload}
\begin{figure}[htpb!]\vspace{-3mm}
\begin{subfigure}{0.5\columnwidth}\vspace{-0mm}
    \centering
    \includegraphics[width=\linewidth, trim=0.01cm 0.01cm 0.01cm 0.01cm, clip=true]{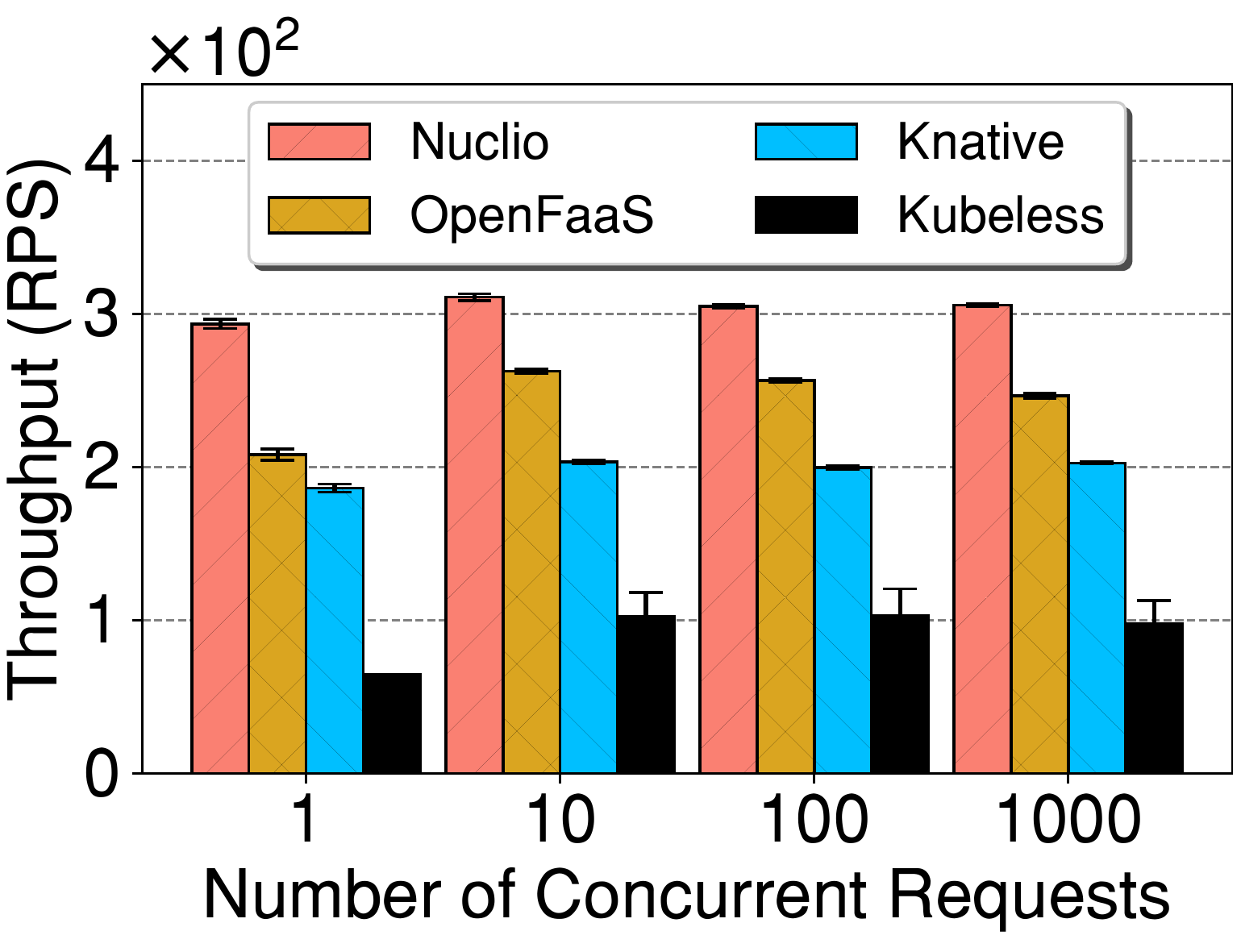}
    \vspace{-5mm}
    \caption{Avg. throughput.}\Scut{for different concurrency levels.}
    \label{fig:HTTPRPS}
\end{subfigure}%
\begin{subfigure}{0.5\columnwidth}\vspace{-0mm}
	\centering 
	\vspace{1mm}
	\includegraphics[width=\linewidth, trim=0.01cm 0.01cm 0.01cm 0.01cm, clip=true]{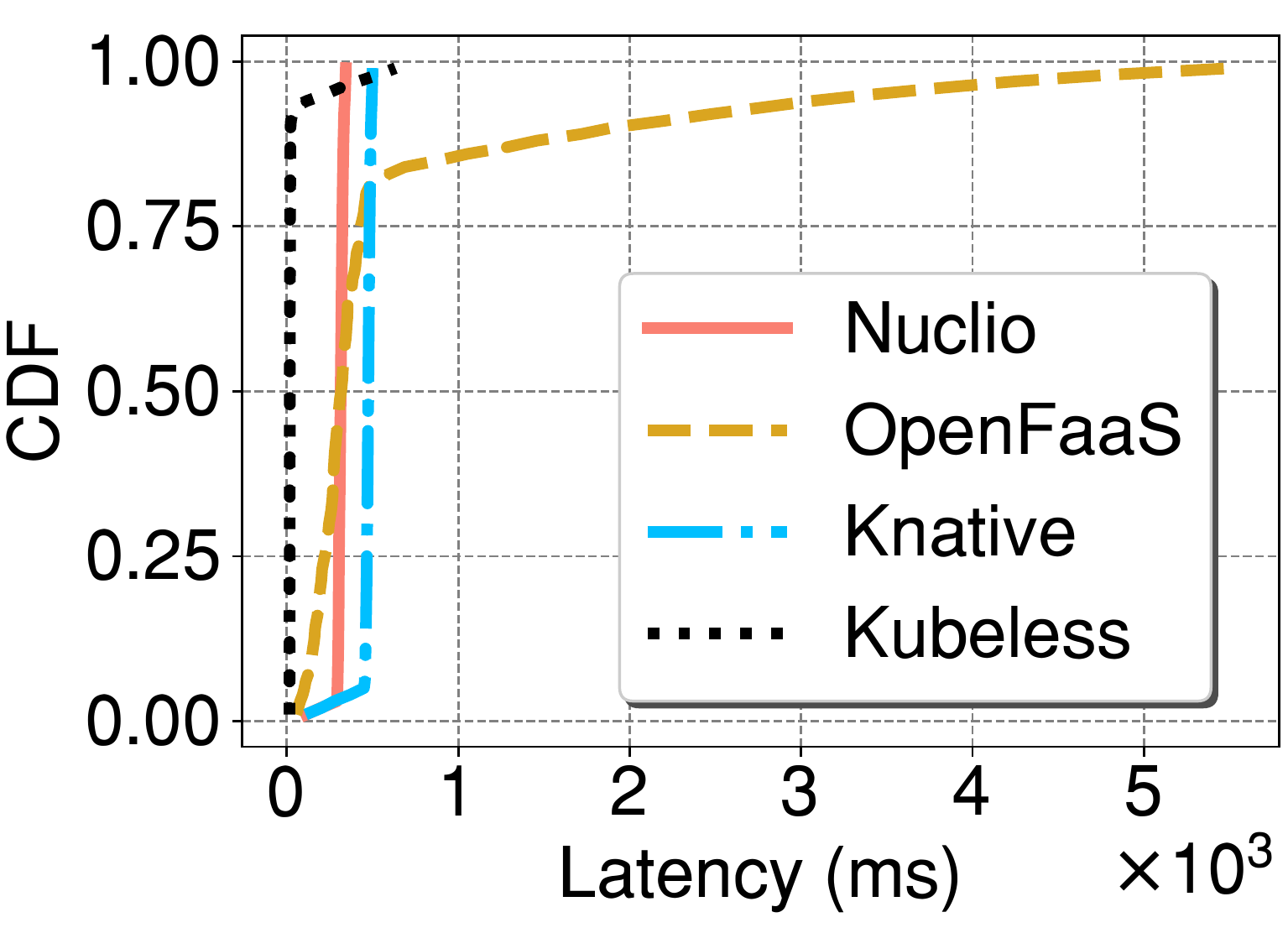}
    \vspace{-5mm}
    \caption{Latency for concurrency of 100.}
    \label{fig:HTTPCDFLatency}
\end{subfigure}%
\vspace{-4mm}
\caption{Throughput \& latency of `http-workload' function across different serverless platforms at different concurrency levels.} 
\label{fig:HTTPWorkload}
\vspace{-4mm}
\end{figure}

\vspace{-1mm}
Next, we change to having http-workload.
Again, we keep the serverless platform settings the same as described in the baseline experiment \S\ref{sec:basline_eval}.
Fig.~\ref{fig:HTTPWorkload} shows the throughput for varying number of concurrent connections and the latency profile for concurrency level of 100. 
Nuclio has the least 99\%ile latency within 500ms,
as it allows queuing only within the function pod, while OpenFaaS and Knative can queue requests at ingress/gateway components. OpenFaaS shows heavy tail due to queuing at both the gateway and watchdog components, each having distinct timeout parameters.
Kubeless drops the connections at the ingress, resulting in additional retries from the client - hence it's lower throughput (the lower latency with Kubeless is because it is measured only for those requests that succeed at a concurrency level of 100).

\vspace{-2mm}\subsubsection{Variable Payload Size}
\begin{figure}[htpb!]\vspace{-3mm}
\begin{subfigure}{0.5\columnwidth}\vspace{-0mm}
    \centering
    \includegraphics[width=\linewidth, trim=0.01cm 0.01cm 0.01cm 0.01cm, clip=true]{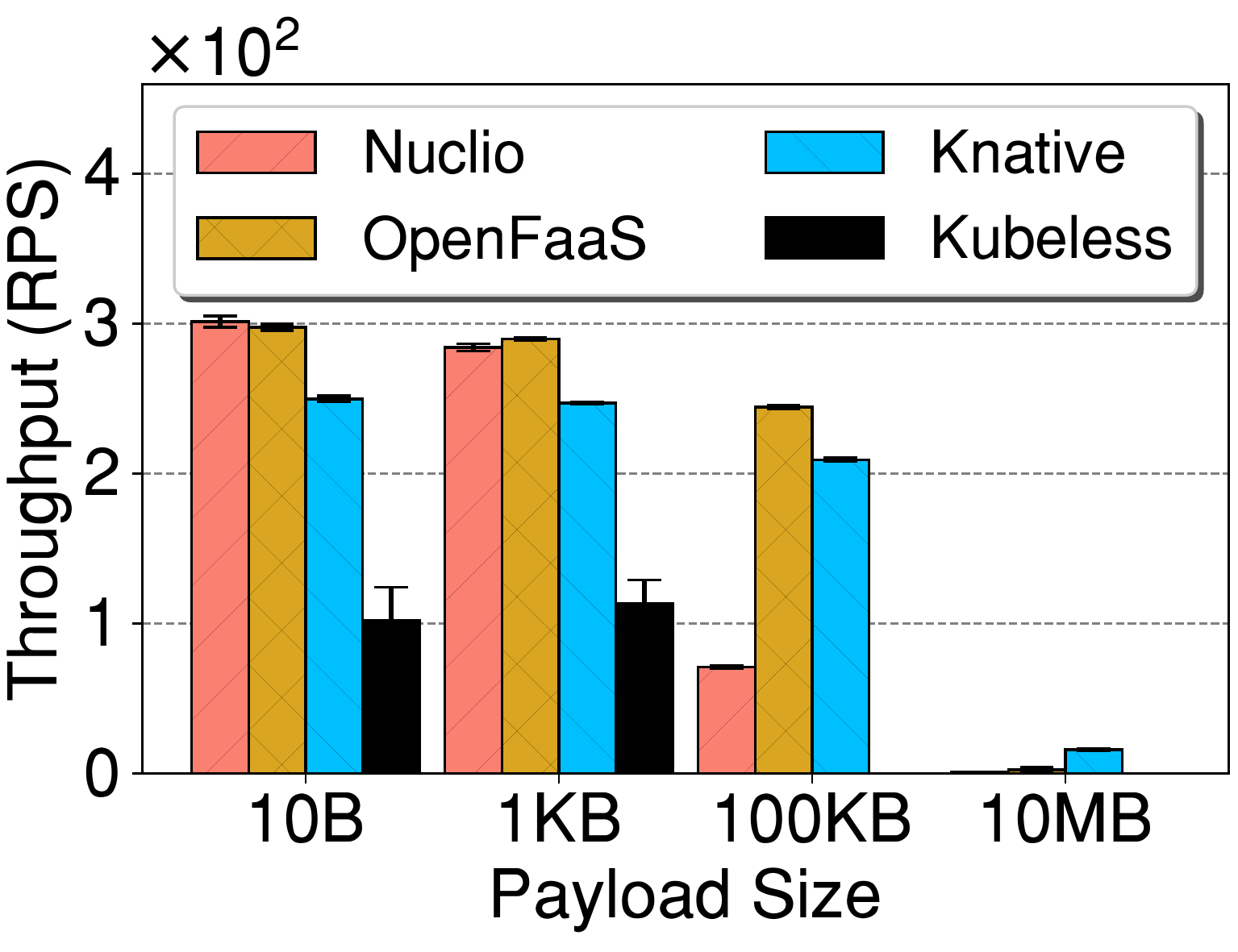}
    \vspace{-5mm}
    \caption{Throughput in requests/second.}
    \label{fig:ThroughputPayload}
\end{subfigure}%
\begin{subfigure}{0.5\columnwidth}\vspace{-0mm}
	\centering
	\vspace{1mm}
	\includegraphics[width=\linewidth, trim=0.01cm 0.01cm 0.01cm 0.01cm, clip=true]{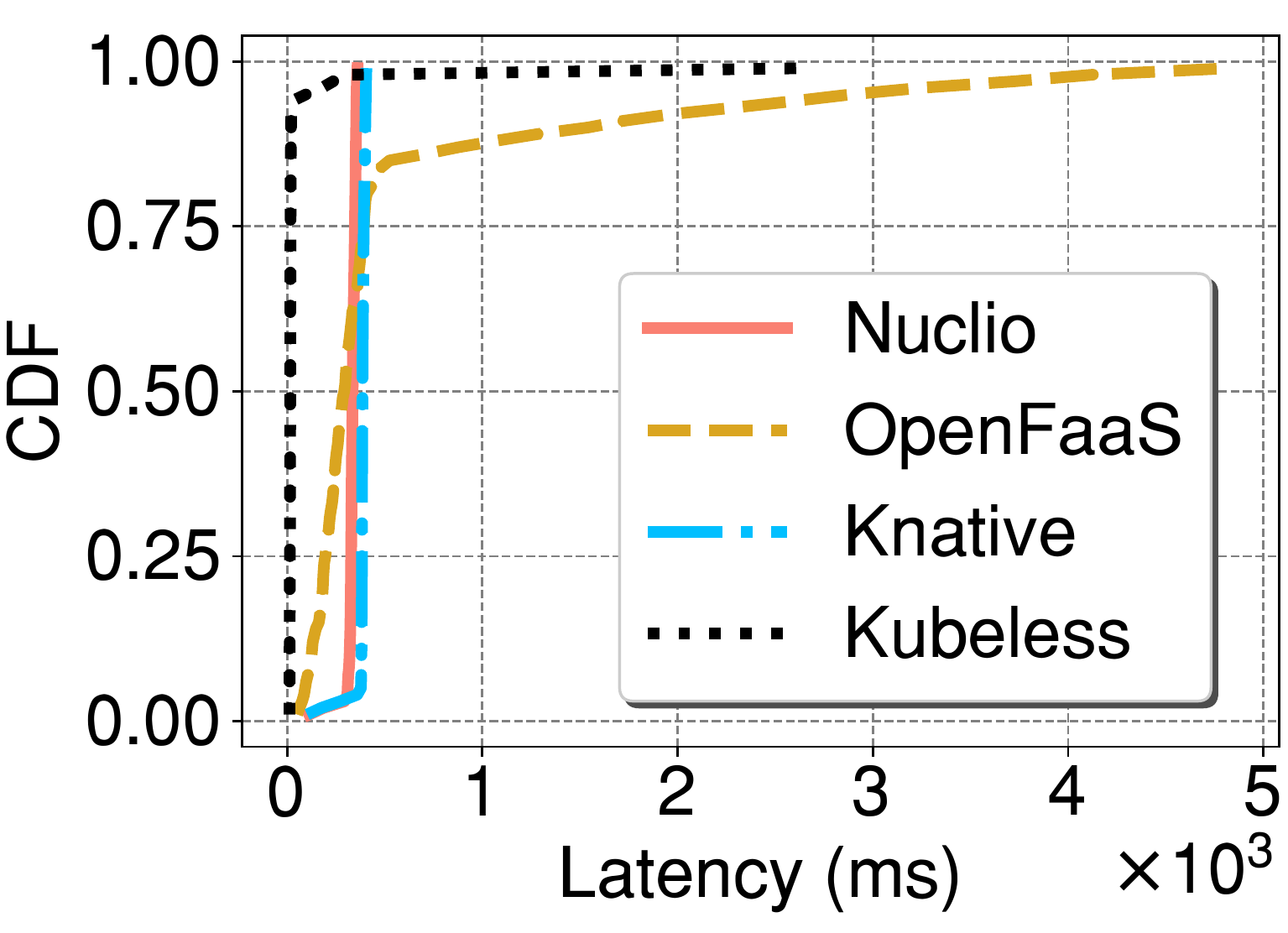}
    \vspace{-5mm}
    \caption{Latency for 1KB payload.}
    \label{fig:LatencyPayload}
\end{subfigure}%
\vspace{-3mm}
\caption{Throughput \& latency of `http-workload' function with different payload sizes for different serverless platforms.}
\label{fig:DiffPayload}
\vspace{-4mm}
\end{figure}

\vspace{-1mm}
For this experiment, in order to assess the data transfer overhead of serverless platforms, we scale the size of payload in the HTTP response and analyze the overheads and impact of assembling, packaging and transporting the HTTP response payload across different serverless platforms.
In Fig.~\ref{fig:DiffPayload}, we observe that Nuclio performs better for small payload sizes (\ie less than 1KB), while OpenFaaS and Knative perform better for large payloads.
\knote{Need to explain why?? Yes, indeed! Please at least speculate}
\knote{additionally, I do not see any insights up to here based on any of the Wireshark trace analysis you folks did. Is there anything you can say with regard to what you learned from the Wireshark trace analysis??}
\vspace{-2mm}

\subsubsection{Impact of different modes of exporting services}
\begin{figure}[htpb!]
\begin{subfigure}{0.5\columnwidth}\vspace{-3mm}
    \centering
    \includegraphics[width=\linewidth, trim=0.01cm 0.01cm 0.01cm 0.01cm, clip=true]{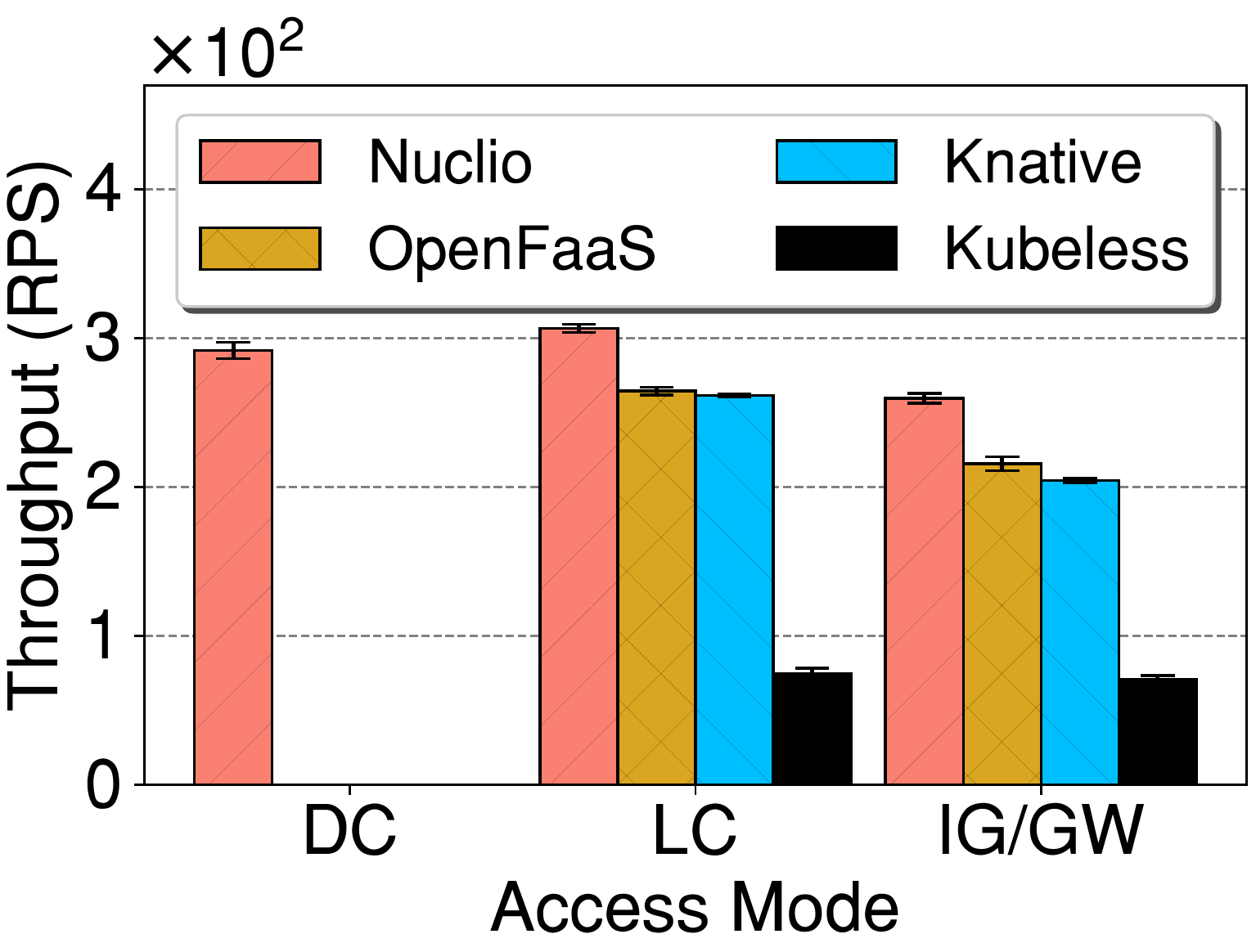}
    \vspace{-6mm}
    \caption{Throughput in requests/second.}
    \label{fig:es:ThroughputDiffModes}
\end{subfigure}%
\begin{subfigure}{0.5\columnwidth}\vspace{-3mm}
	\centering
	\vspace{1mm}
	\includegraphics[width=\linewidth, trim=0.01cm 0.01cm 0.01cm 0.01cm, clip=true]{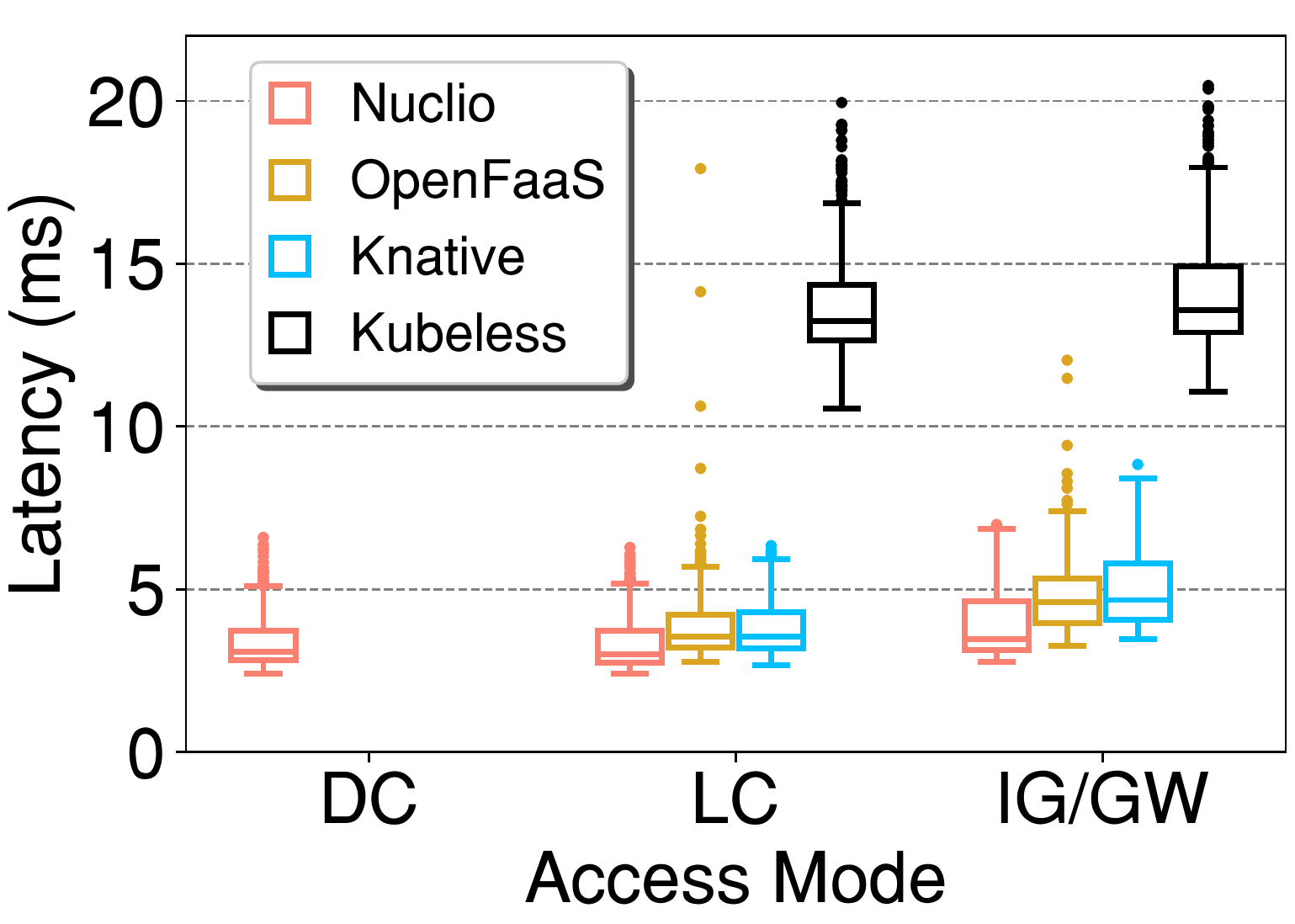}
    \vspace{-6mm}
    \caption{Latency in ms.}
    \label{fig:es:LatencyDiffModes}
\end{subfigure}%
\vspace{-4mm}
\caption{Throughput \& latency for different methods of exporting the services on different serverless platforms.}
\label{fig:es:DiffModes}
\vspace{-4mm}
\end{figure}

\vspace{-1mm}
In order to avoid the added queuing latency, we run the http workload with \texttt{wrk} tool and limit the number of maximum concurrent (in-flight) requests to 1 and repeat the experiment 1000 times. 
Fig.~\ref{fig:es:DiffModes} shows the impact on throughput and latency for three different modes of exporting and invoking the serverless functions. 
\textbf{LC} refers to local call, where the client and function pods reside on the same node in the Kubernetes cluster, and client invokes the function directly using the IP-address of the function pod. Nuclio has marginally better throughput and lower latency than Knative and OpenFaaS, while Kubeless suffers in both latency and throughput. 
\textbf{IG/GW} refers to exporting and invocation of serverless function through the ingress/API gateway components. This mode brings down the throughput across all platforms, and also incurs ($\sim$ 1ms) additional latency than `LC' mode.
\textbf{DC} (direct call) approach is only supported by Nuclio, which exports the function pod using the nodeport service of Kubernetes. DC 
avoids the additional routing overheads in the worker node (netfilter rules that translate the packet destination, and forward the packets to the function pod). 

\vspace{-1mm}\subsubsection{Analysing the latency impact of serverless platforms}
\begin{figure}[!htb]\vspace{-3mm}
    \centering
    \begin{minipage}{0.34\columnwidth}
    \vspace{0pt}
        \centering
        \includegraphics[width=0.7\linewidth, trim=0.01cm 0.01cm 0.01cm 0.01cm, clip=true]{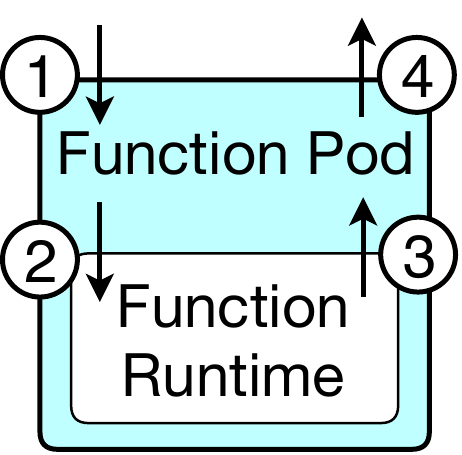}
        \vspace{-1mm}
        \label{fig:LatencyBD-1}
    \end{minipage}
    \begin{minipage}{.65\columnwidth}
        \centering
        \captionsetup{type=table} 
        \vspace{0pt}
        \begin{tabular}{c|ccc}
        \hline
        Process & 1$\rightarrow$2 & 2$\rightarrow$3 & 3$\rightarrow$4 \\
        \hline
        Nuclio & 0.63 & 0.001 & 0.54 \\
        OpenFaaS & 1.32 & 0.001 & 0.93 \\
        Knative & 1.30 & 0.001 & 0.62 \\
        Kubeless & 4.96 & 0.001 & 2.63 \\
        \hline
        \end{tabular}
        \vspace{-1mm}
        \label{table:LatencyBD-2}
    \end{minipage}
    \vspace{-3mm}
    \caption{Latency breakdown (ms) parts of serverless execution.}
    \label{fig:LatencyBD}
    \vspace{-4mm}
\end{figure}

\vspace{-1mm} 
We analyze the delay overheads incurred in processing the serverless functions for different platforms. We breakdown the processing delays within the function pod. For this experiment, we use \texttt{curl} to send one request for `hello-world' function and use \texttt{tcpdump} to capture the packets on the worker node of the function pod. We record four timestamps,
\ie (1) when the request reaches the function pod; (2) start of the function runtime; (3) end of the function runtime; (4) when the response is sent out of function pod.  
Time intervals between these timestamps are shown in Fig.~\ref{fig:LatencyBD}.
In all frameworks, the actual run-time of the function (0.001ms) is the same. However, the function initiation time (time taken for request to be forwarded to the function instance) and function response delay (time taken for the response of the function to be sent out of the pod) vary. This depends on how the data is packaged and shared with the function instance. Also, Kubeless (due to forking per request), incurs very high delay in forwarding the packet to the function instance. We also experimented with the 'http-function' and found the startup and response delay overheads to be same.

\vspace{-1mm}\subsection{Auto-scaling}\label{eval:autoscaling}\vspace{-1mm}
Auto-scaling capabilities exported by different serverless platforms vary. Here, we compare the auto-scaling features of Knative and OpenFaaS for both the rate-based and Kubernetes-based horizontal-pod-autoscaler (HPA) modes under different workload characteristics. 
For a fair comparison, we tune the auto-scaling related configuration parameters in both the platforms to have the same interval for the auto-scale triggers and factors for scaling functions.\footnote{In Knative, we disable panic mode, and set the minScale and maxScale instances as 1 and 10, target to 10, max-scale-up-rate to 100, tick interval to 2s, and stable window to 10s, which ensures triggering auto-scale notifications on a 2s window and scaling to 1 or more instances at a time. Likewise, for OpenFaaS, we set scale-factor to 10 and configure the alert-notification window to 2s, and RPS threshold to 10. For HPA, we set CPU limits to 50.} We use the same python function as in \S\ref{sec:httpworkload}.
\noindent\textbf{Note:} In OpenFaaS, auto-scaling is based on the average rate of the incoming requests 
(RPS);
in Knative, auto-scaling is based on the concurrency level observed per function instance. The subtle difference is that the average RPS value can be lower or higher than the observed concurrency depending on whether the time for processing a function invocation is higher or lower respectively. We will demonstrate the benefit and deficiency of both approaches. \hide{say autoscale_threshold qps=10, c=10 for scaling, case 1: if workload has rps=11 unbounded traffic, OF will scale, KN=?
case 2: processing cost = 1ms then OF and KN will both scale while 
case 3: processing cost = 1sec
case 4: 
}
\vspace{-1mm}\subsubsection{Workload based auto-scaling}
\begin{figure}[htb!]
\begin{subfigure}{0.5\columnwidth}\vspace{-3mm}
    \centering
    \includegraphics[width=\linewidth, trim=0.01cm 0.01cm 0.01cm 0.01cm, clip=true]{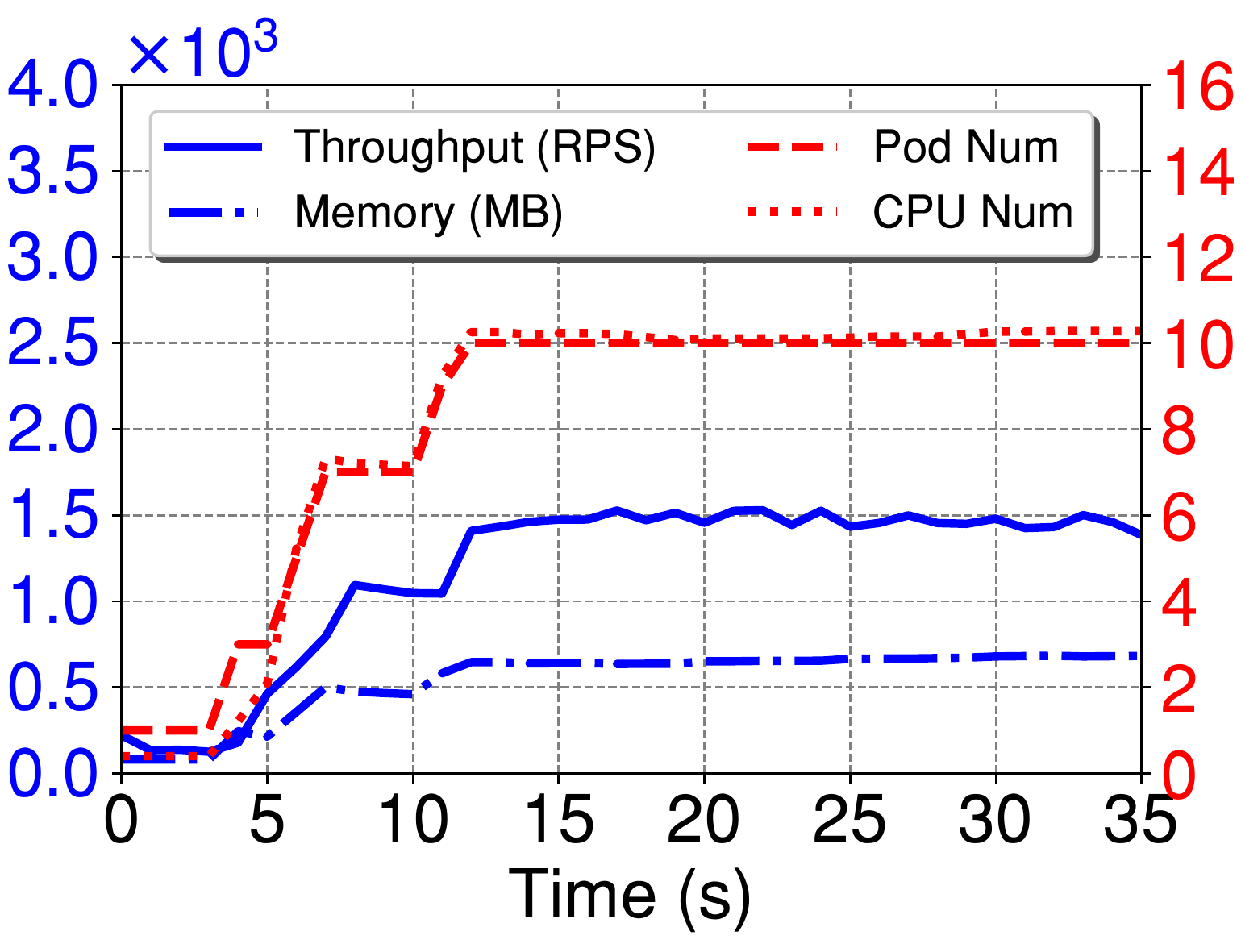}
    \vspace{-5mm}
    \caption{Knative.}
    \label{fig:Autoscaling-qps-steady-knative}
\end{subfigure}%
\begin{subfigure}{0.5\columnwidth}\vspace{-3mm}
	\centering
	\includegraphics[width=\linewidth, trim=0.01cm 0.01cm 0.01cm 0.01cm, clip=true]{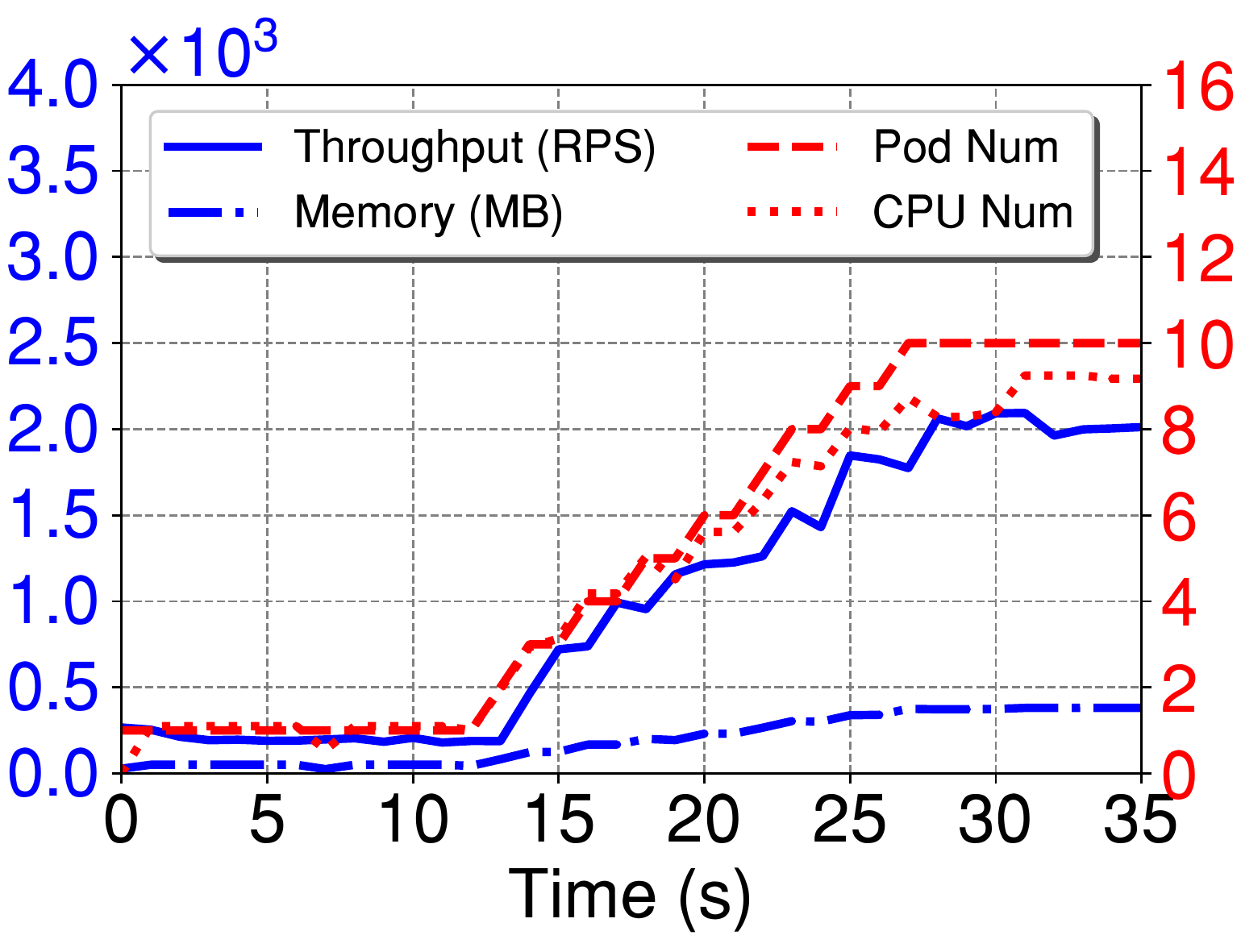}
    \vspace{-5mm}
    \caption{OpenFaaS.}
    \label{fig:Autoscaling-qps-steady-openfaas}
\end{subfigure}%
\vspace{-3mm}
\caption{Auto-scaling with steady workload.}
\label{fig:Autoscaling-qps-steady}
\end{figure}

\vspace{-1mm}
\textbf{Steady workload}:
We use the \texttt{wrk} tool, set a steady rate for outstanding requests (concurrency of 100) and run the experiment for 60s.
Also, to enforce proper traffic distribution across newly created pods, we force the OpenFaaS gateway to terminate and reestablish connections with the function pods.
Periodically, every 2s, we monitor the number of pod instances, CPU and memory usage, and throughput.
From Fig.~\ref{fig:Autoscaling-qps-steady}, we observe that Knative scales multiple instances at a time to reach the max. (10) instances quickly (in 12s), 
while OpenFaaS just scales up one instance at a time, taking 26s to scale up to the max. (10) instances. 
Although, the CPU usage for the scaled instances looks identical, the memory pressure is higher for Knative. This stems from the difference in the python runtimes and overheads in the queue-proxy container component for Knative and of-watchdog components in OpenFaaS.
  
\vspace{-4mm}



\begin{figure}[htb!]
\begin{subfigure}{0.5\columnwidth}\vspace{-3mm}
    \centering
    \includegraphics[width=\linewidth, trim=0.01cm 0.01cm 0.01cm 0.01cm, clip=true]{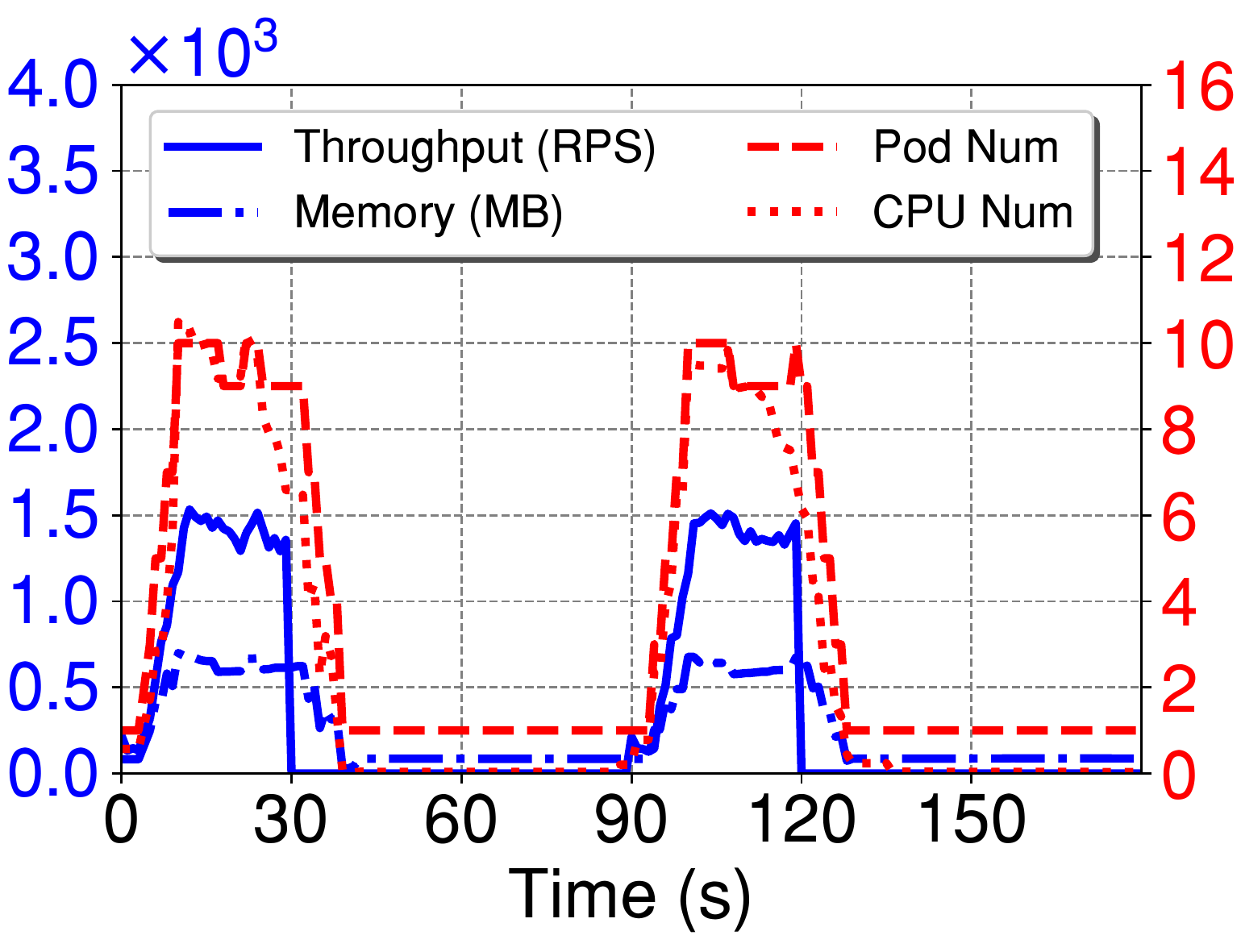}
    \vspace{-5mm}
    \caption{Knative.}
    \label{fig:Autoscaling-qps-bursty-knative}
\end{subfigure}%
\begin{subfigure}{0.5\columnwidth}\vspace{-3mm}
	\centering
	\includegraphics[width=\linewidth, trim=0.01cm 0.01cm 0.01cm 0.01cm, clip=true]{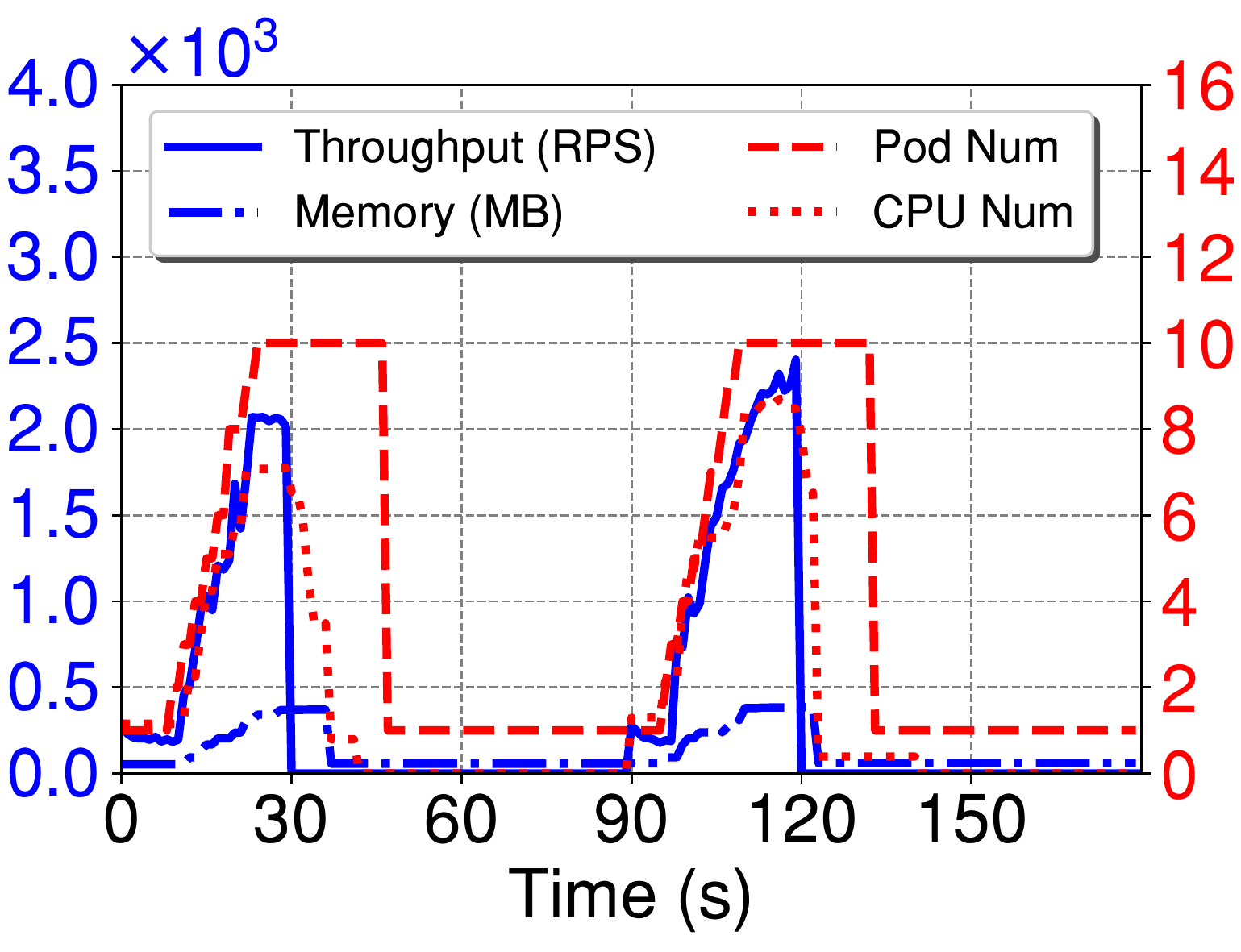}
    \vspace{-5mm}
    \caption{OpenFaaS.}
    \label{fig:Autoscaling-qps-bursty-openfaas}
\end{subfigure}%
\vspace{-3mm}
\caption{Auto-scaling with bursty workload.}
\label{fig:Autoscaling-qps-bursty}
\end{figure}

\noindent \textbf{Bursty workload}
We also experimented by varying the http workload to have bursts of concurrent requests 
followed by a large idle period. Fig.~\ref{fig:Autoscaling-qps-bursty} shows that, Knative is more responsive to bursts, and is able to scale quickly to a large number of instances, 
while OpenFaaS scales 
gradually 
and has lower average throughput.


\vspace{-4mm}

\begin{figure}[htb!]\vspace{-3mm}
\begin{subfigure}{0.5\columnwidth}\vspace{-3mm}
    \centering
    \includegraphics[width=\linewidth, trim=0.01cm 0.01cm 0.01cm 0.01cm, clip=true]{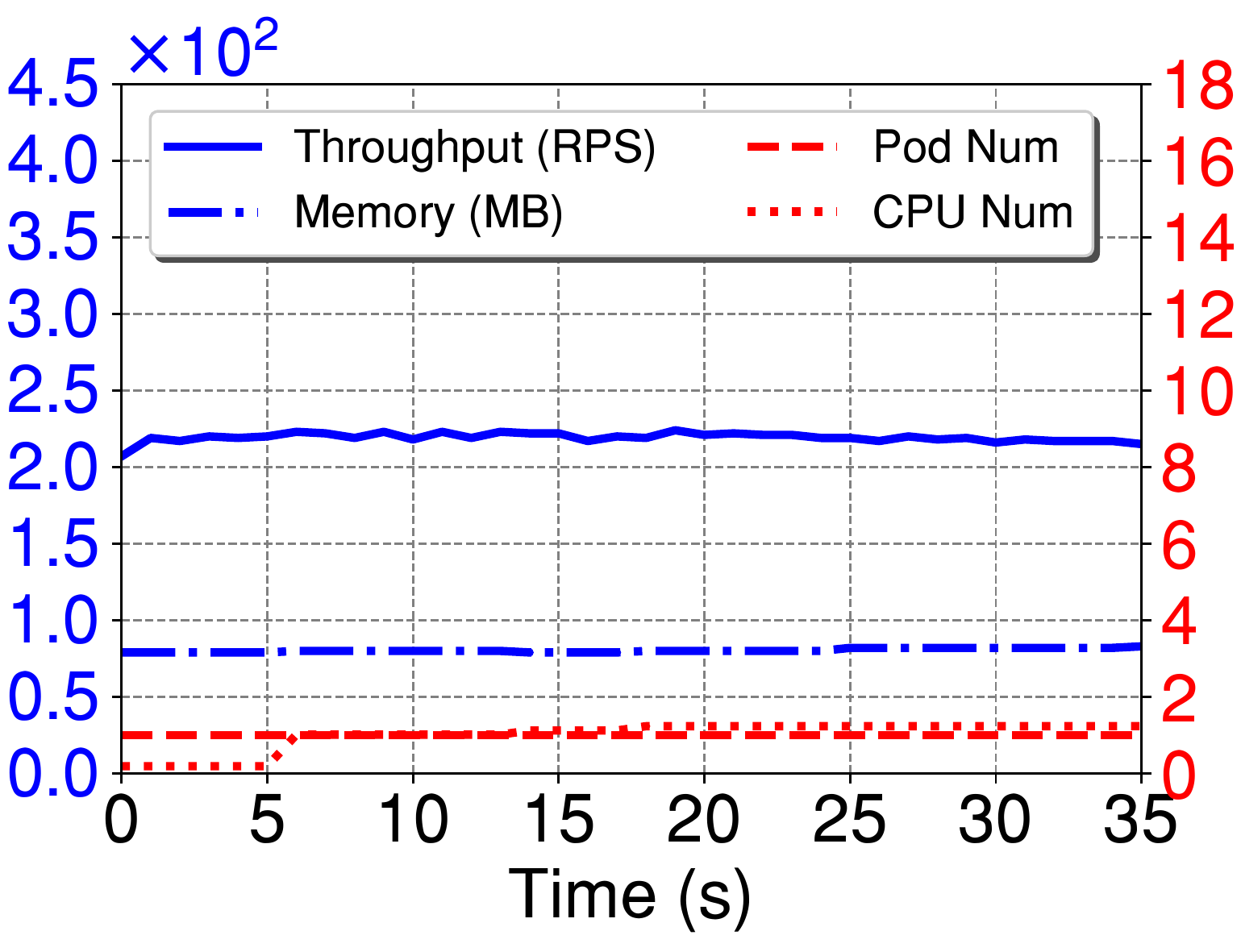}
    \vspace{-5mm}
    \caption{Knative.}
    \label{fig:Autoscaling-issue-knative}
\end{subfigure}%
\begin{subfigure}{0.5\columnwidth}\vspace{-2mm}
	\centering
	\includegraphics[width=\linewidth, trim=0.01cm 0.01cm 0.01cm 0.01cm, clip=true]{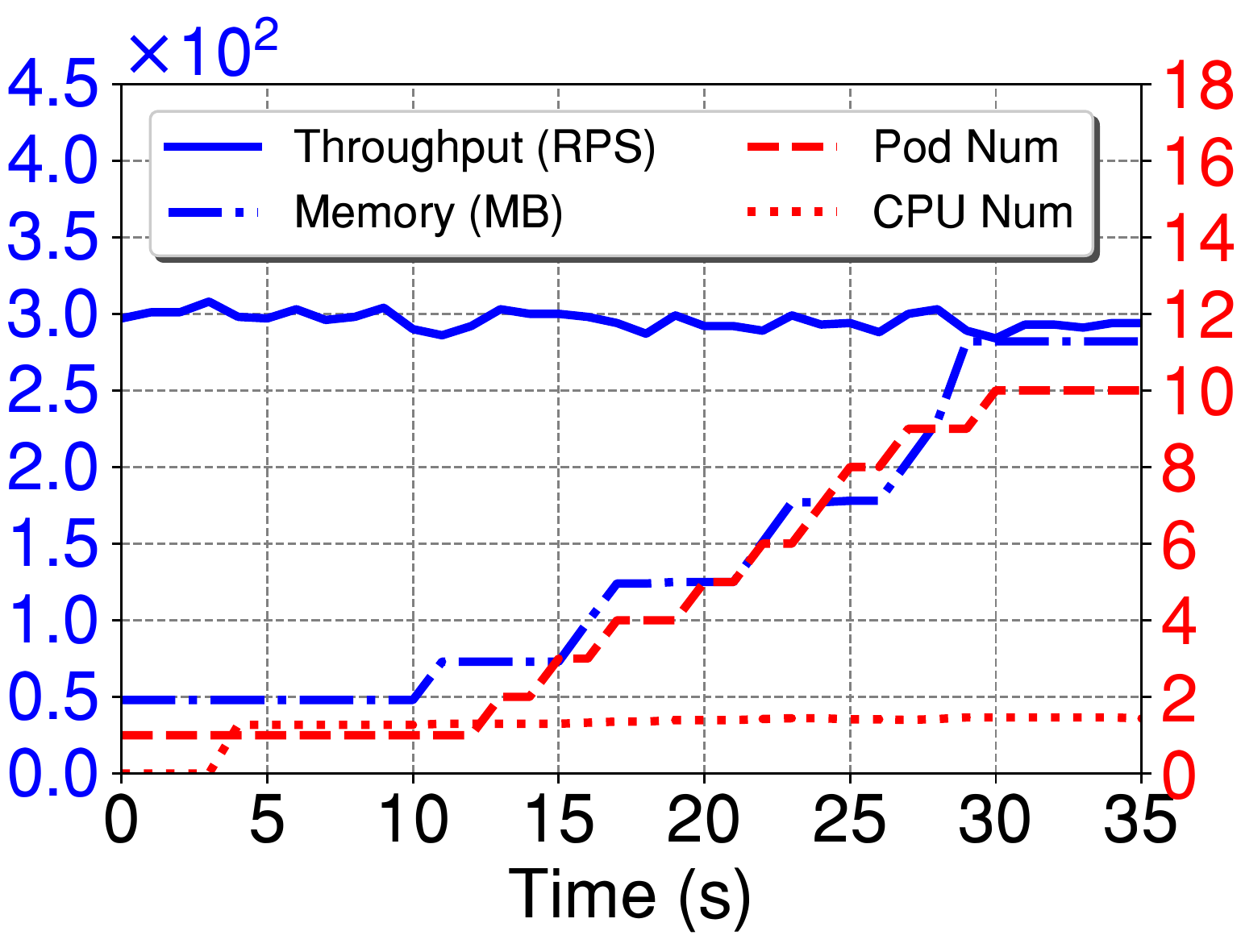}
    \vspace{-5mm}
    \caption{OpenFaaS.}
    \label{fig:Autoscaling-issue-openfaas}
\end{subfigure}%
\vspace{-3mm}
\caption{Auto-scaling issues with Knative and OpenFaaS.}
\label{fig:Autoscaling-issue}
\vspace{-4mm}
\end{figure}

\noindent\textbf{Issues with auto-scaling in OpenFaaS and Knative}:
In another experiment, we use the same setup as in the steady workload 
experiment, but lower the number of outstanding requests from 100 to 9. 
From Fig.~\ref{fig:Autoscaling-issue}, we observe that Knative fails to auto-scale and continues to operate with just 1 function pod instance, resulting in almost 7$\times$ lower (200 RPS) throughput compared to the earlier case (1500 RPS).
Next, we revert to vanilla OpenFaaS (\ie as in github, and disable the workaround of resetting the connections to the function pods), and run the same steady workload experiment. The function pods get auto-scaled as before. But, the throughput shows no improvement. Also, note that with auto-scaling the memory usage increases, but CPU utilization remains steady. We found the issue to be due to incorrect traffic distribution. Due to the long running connections (setup by OpenFaaS gateway at the beginning with the first function pod), all the traffic is just routed only to the first function pod, while the remaining, newly scaled pods, do not receive any traffic.\footnote{
Bug raised: \url{https://github.com/openfaas/faas/issues/1303}.}

\vspace{-3mm}\subsubsection{Resource based auto-scaling}
\begin{figure}[htb!]\vspace{-1mm}
\begin{subfigure}{0.5\columnwidth}
    \vspace{-2mm}
    \centering
    \includegraphics[width=\linewidth, trim=0.01cm 0.01cm 0.01cm 0.01cm, clip=true]{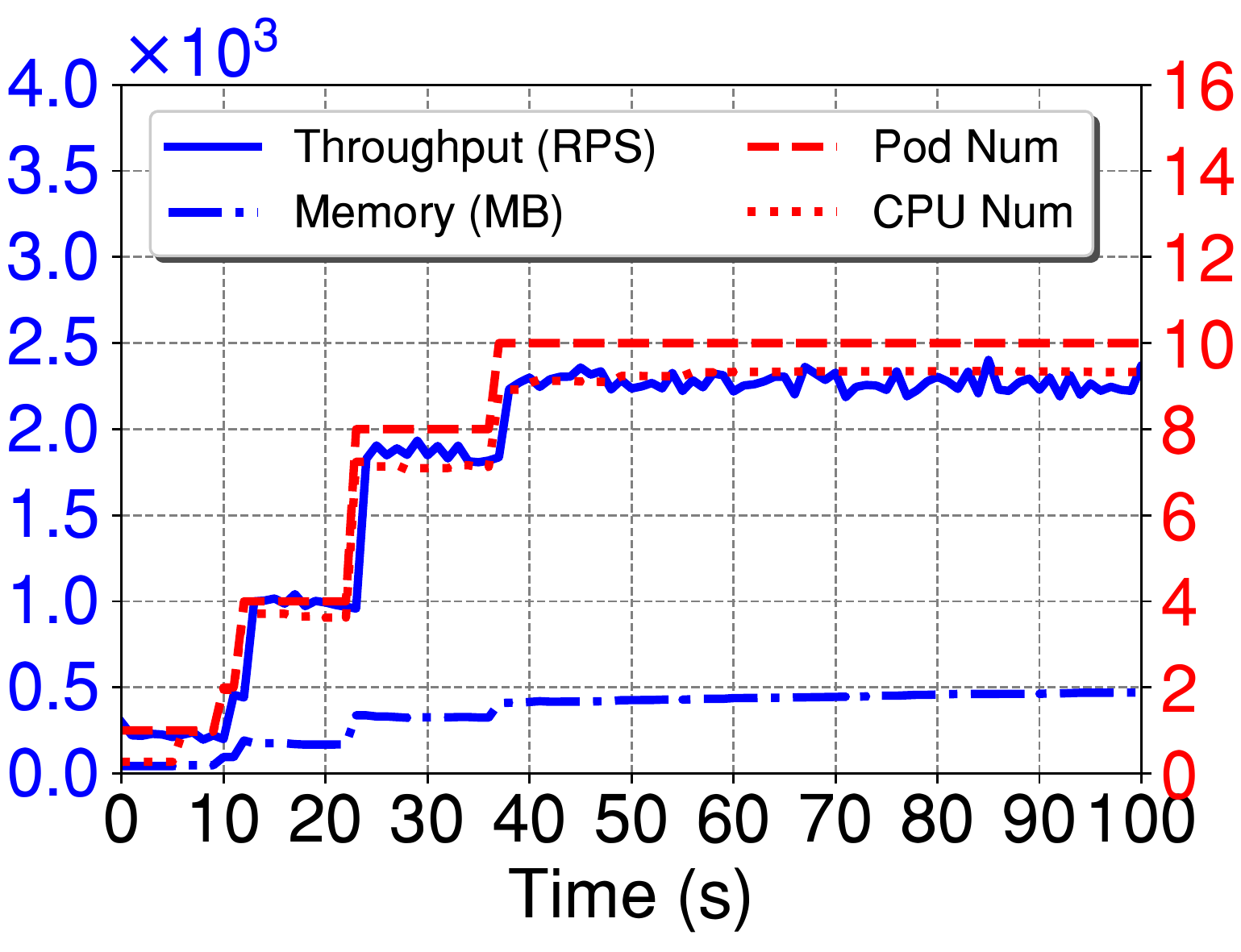}   
    \vspace{-6mm}
    \caption{Nuclio.}
    \label{fig:Autoscaling-hpa-steady-nuclio}
\end{subfigure}%
\begin{subfigure}{0.5\columnwidth}\vspace{-3mm}
	\centering
	\includegraphics[width=\linewidth, trim=0.01cm 0.01cm 0.01cm 0.01cm, clip=true]{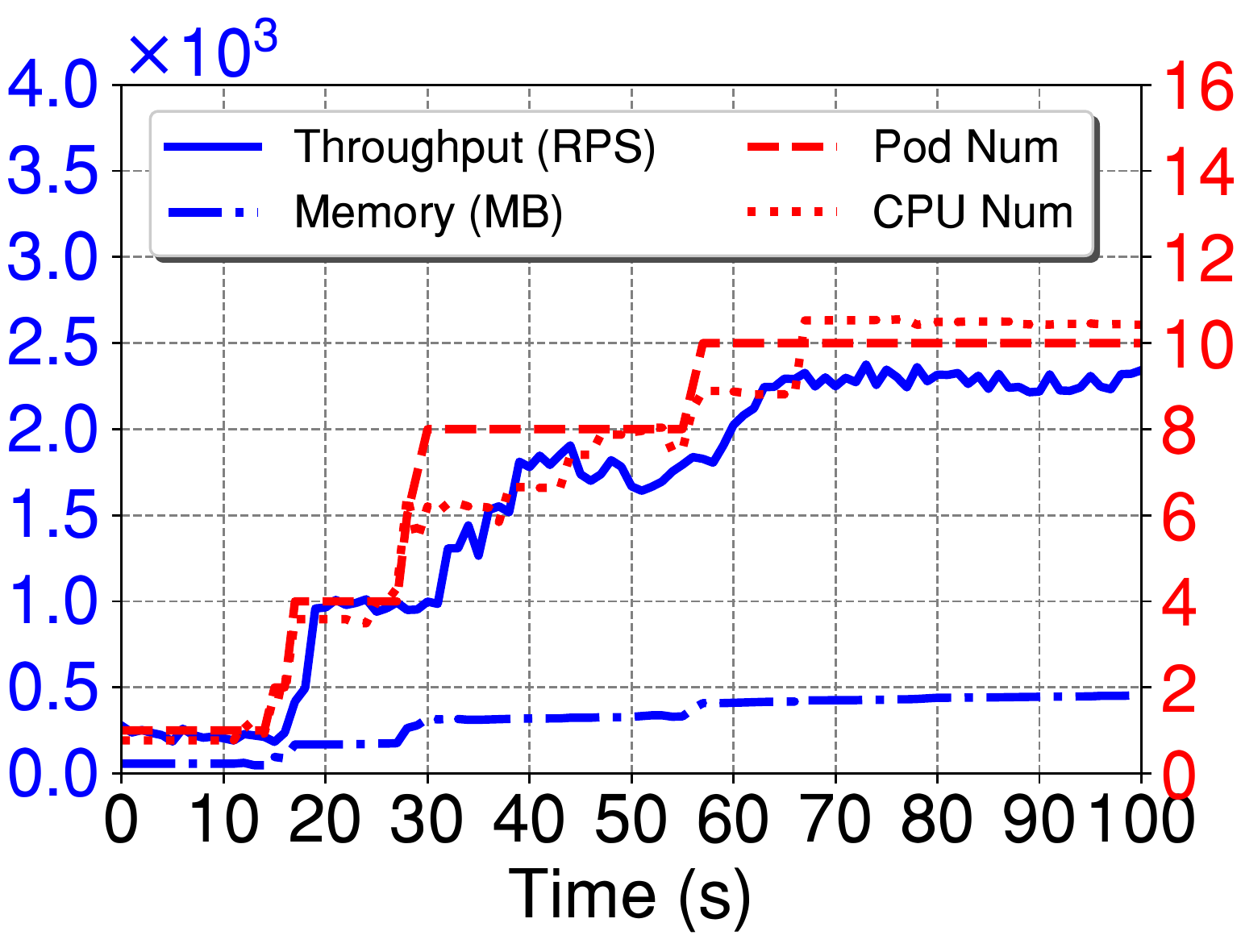}
    \vspace{-6mm}
    \caption{OpenFaaS.}
    \label{fig:Autoscaling-hpa-steady-openfaas}
\end{subfigure}
\begin{subfigure}{0.5\columnwidth}\vspace{-1mm}
    \centering
    \includegraphics[width=\linewidth, trim=0.01cm 0.01cm 0.01cm 0.01cm, clip=true]{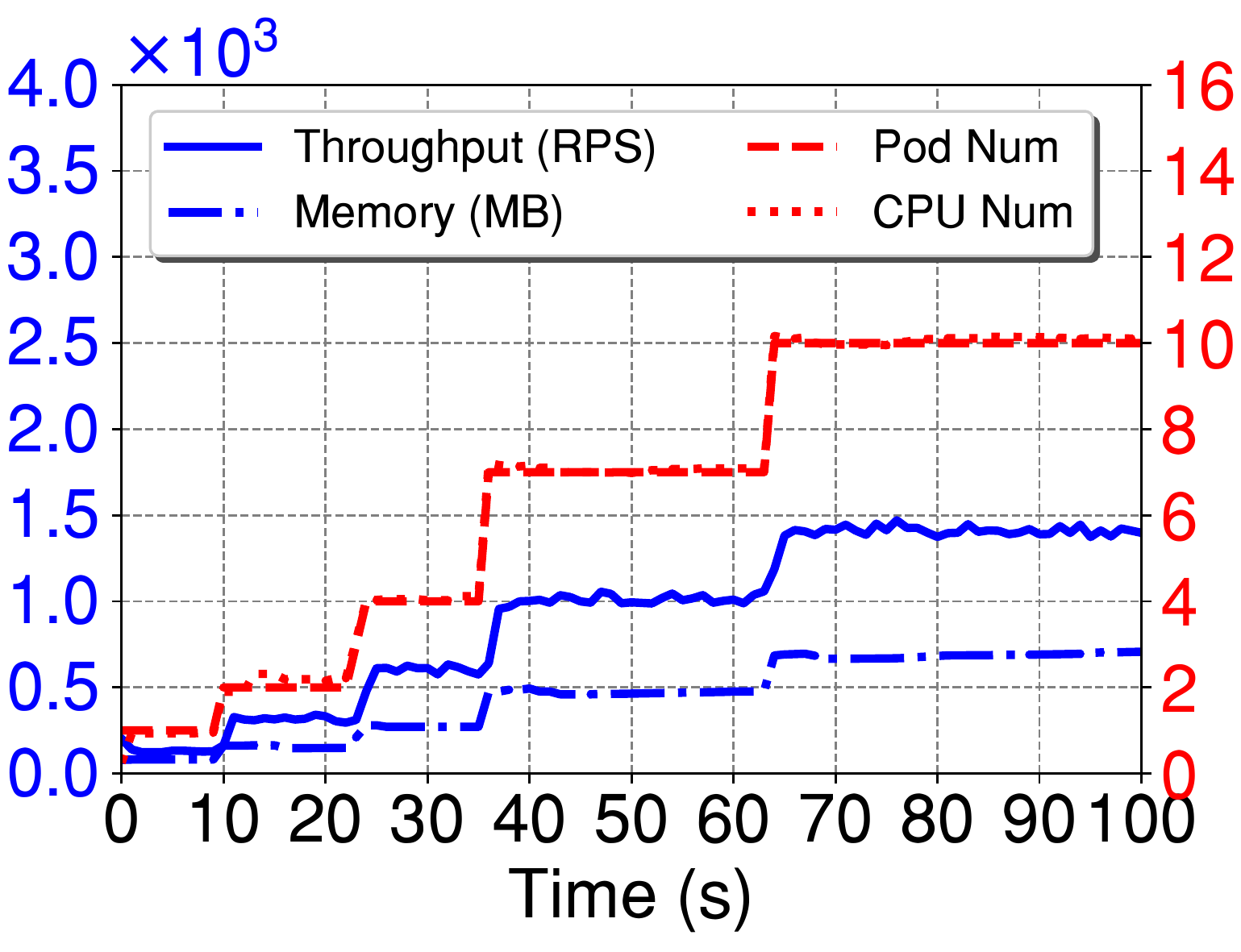}
    \vspace{-6mm}
    \caption{Knative.}
    \label{fig:Autoscaling-hpa-steady-knative}
\end{subfigure}%
\begin{subfigure}{0.5\columnwidth}\vspace{-1mm}
    \centering
    \includegraphics[width=\linewidth, trim=0.01cm 0.01cm 0.01cm 0.01cm, clip=true]{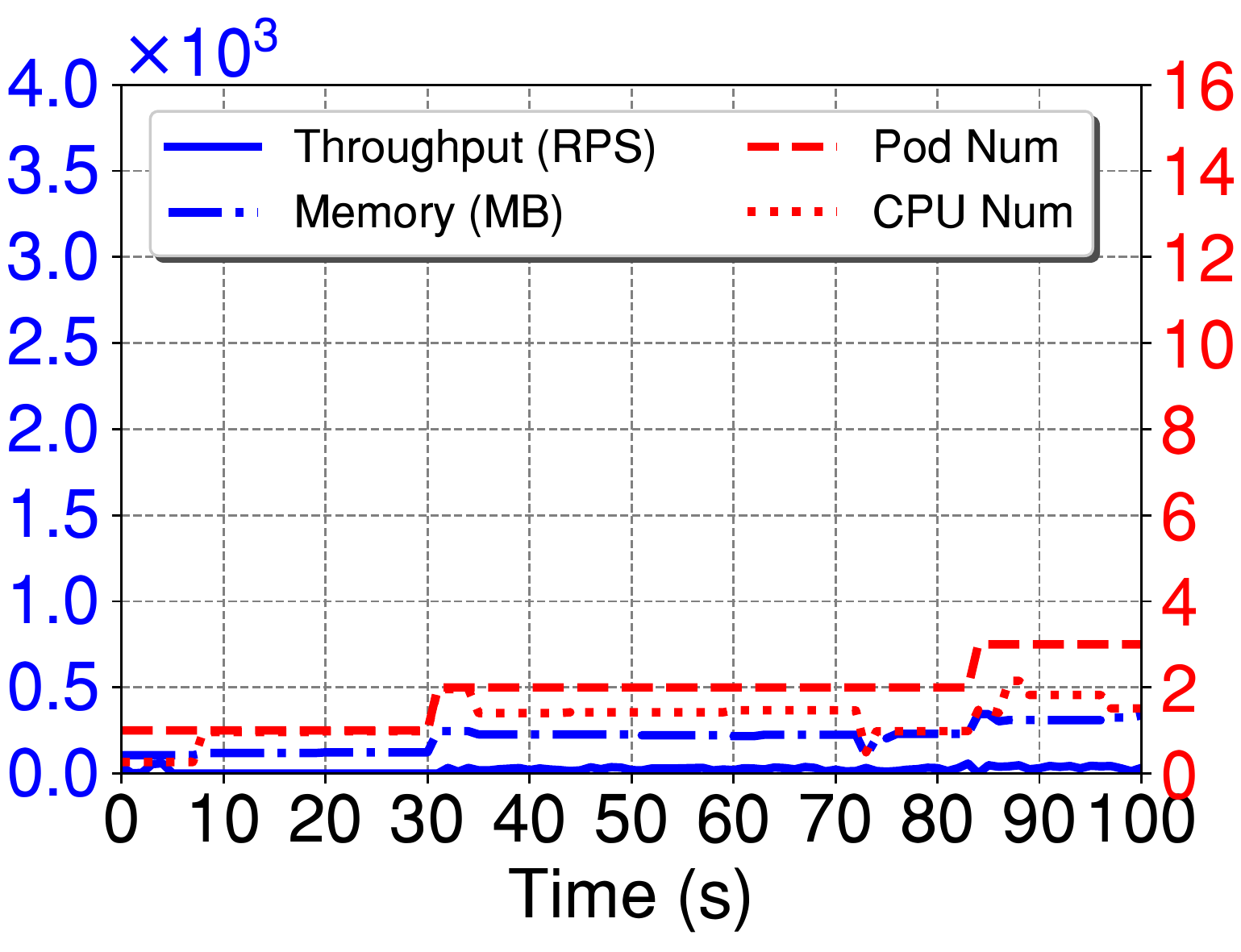}
    \vspace{-6mm}
    \caption{Kubeless.}
    \label{fig:Autoscaling-hpa-steady-kubeless}
\end{subfigure}%
\vspace{-4mm}
\caption{HPA-based auto-scaling on steady workload.}
\label{fig:Autoscaling-hpa-steady}
\vspace{-6mm}
\end{figure}

\vspace{-1mm}
We use the same setup (steady state), configure the cpu usage limit to 50\%, and leverage Kubernetes HPA for auto-scaling. Note, the auto-scaling of function pods is governed by Kubernetes only. 
From Fig.~\ref{fig:Autoscaling-hpa-steady}, we can observe that, except for Kubeless, the auto-scaling behavior is same across all the platforms \ie auto-scaling tries to double the instances at each step until it reaches the maximum (10). However, the duration of each step depends on the CPU utilization factor, which in turn depends on the serverless platform specific components (event-listener, of-watchdog, queue-proxy). Nuclio, being relatively more CPU hungry is able to scale more rapidly (in 40s), than Knative and OpenFaaS. 
With Kubeless, the fork-per-request results in high latency, dropping of incoming requests that in turn results in low throughput and low CPU utilization. Thus, it results in poor auto-scaling as well.

\vspace{-1mm}
\section{Related Work}\vspace{-1mm}
\noindent \textbf{Serverless Platform comparison}: In work~\cite{mcgrath2017serverless, wang2018peeking}, the authors conducted several measurements on different cloud serverless platforms (AWS Lambda, Microsoft Azure, Google Cloud), and found the AWS to be better in terms of throughput, scalability, cold-start latency.  The works~\cite{lloyd2018serverless,cui2} investigate the different factors that influence the performance of AWS lambda, namely the impact of the choice of language of the function, memory footprint of the function, \etc Work~\cite{mohanty2018evaluation} evaluates the performance of Fission, Kubeless and OpenFaaS serverless frameworks and characterizes the response time and the ratio of successfully completed requests for different loads. However the work fails to characterize the throughput of these platforms and accounts for the mean latency (response time) and successful responses at different load characteristics, which is debatable, without the proper consideration and configuration of the serverless platform specific configuration parameters, resulting in inaccurate results. 
In the most recent work~\cite{palade}, the authors quantitatively evaluate  Apache OpenWhisk, OpenFaas, Kubeless, and Knative platforms. The results for Kubeless are similar, but for the other platforms, we feel the presented results are inaccurate. This could be due to the usage of Kubernetes. 
In contrast, our work focuses on discerning the architectural blocks that impact the performance of Kubernetes based open-source serverless platforms.


\vspace{-2mm}
\section{Summary}\vspace{-1mm}
Through measurements, we explored different open-source 
serverless platforms and identified the key design considerations and their impact on performance and auto-scaling. We show that the interaction between the API Gateway/Ingress controller and the function pods, the overheads of this component and the way requests are queued influence baseline performance. 
Further, the `RPS'-based and `Concurrency'-based auto-scaling approaches by themselves are insufficient and need to evolve to properly meet workload demands, so that we can avoid maintaining a large number of instances active. 
\noindent {\textbf{Acknowledgements}:}
This work was supported by US NSF grants CRI-1823270 and CNS-1763929, and grants from Hewlett Packard Enterprise Co., Futurewei Technologies Inc, and the National Key Research and Development Program of China under Grant 2018YFB1800100, 2018YFB1800500, 2018YFB1800800, and China Scholarship Council.

\vspace{-3mm}
\bibliographystyle{ACM-Reference-Format}
\bibliography{src/sigproc_old}





\end{document}